\documentclass[11pt,a4paper]{article}
\usepackage{jheppub}

\pdfoutput=1

\usepackage{tabularx}
\usepackage{amsmath,amssymb}
\usepackage{appendix}
\usepackage{graphicx}
\usepackage{color}
\usepackage{cancel}
\usepackage{comment}

\newcolumntype{L}[1]{>{\hsize=#1\hsize\raggedright\arraybhttps://www.overleaf.com/project/5fb496471b52212baf93df33ackslash}X}%
\newcolumntype{R}[1]{>{\hsize=#1\hsize\raggedleft\arraybackslash}X}%
\newcolumntype{C}[2]{>{\hsize=#1\hsize\columncolor{#2}\centering\arraybackslash}X}%

\preprint{IPMU22-0030}

\title{Generating Non-topological Solitons Via Thermal Corrections: Higgs Balls}
\author[a]{Lauren Pearce}
\affiliation[a]{Pennsylvania State University-New Kensington, New Kensington, PA 15068}
\emailAdd{lpearce@psu.edu}

\author[b]{Graham White}
\affiliation[b]{Kavli IPMU (WPI), UTIAS, The University of Tokyo, Kashiwa, Chiba 277-8583, Japan}
\emailAdd{graham.white@ipmu.jp}

\author[c,b]{Alexander Kusenko}
\affiliation[c]{Department of Physics and Astronomy, UCLA, Los Angeles, CA 90095, USA}
\emailAdd{kusenko@g.ucla.edu}

\abstract{Scalar fields which carry charge can generally form non-topoligical solitons (Q-balls), if the energy in the extended configuration is less than the energy of an equivalent number of free quanta.  For global Q-balls, such solitons exist whenever the potential grows slower than quadratically.  We show that even in the absence of attractive interactions, finite temperature corrections can generate Q-ball solutions, as the coefficient of cubic corrections is generally negative.  As an illustration of this, we consider the possibility of constructing Q-balls using the Higgs field.  We first show that the finite temperature corrections would enable the existence of Higgs balls if the Standard Model symmetry was ungauged.  We then show that Higgs self-interactions mediated by the Standard Model gauge bosons are sufficient to prevent the existence of these states in the actual Standard Model, but can be present in a variety of extensions.}
\begin{document}

\maketitle

\section{Introduction}
\label{intro}

\paragraph{}Non-topological solitons, or Q-balls, are extended configurations of a scalar field whose stability is typically ensured by a conserved charge.  Self-interactions ensure that the energy of the extended configuration with charge $Q$ is less than the energy of $Q$ quanta of the scalar field~\cite{Rosen:1968mfz,Friedberg:1976me,Coleman:1985ki,Kusenko:1997ad}.  For models with a global charge, this occurs when $\sqrt{ V(\phi) \slash \phi^2}$ is minimized at nonzero VEV $\phi$~\cite{Coleman:1985ki}.  

\paragraph{}In this work, we first observe that in a wide range of models, finite temperature corrections can allow Q-balls which do not existence at zero temperature.  (This is different than studying thermal corrections in systems which admit Q-balls at zero temperature, as in e.g.~\cite{Laine:1998rg}.)  In particular, symmetry principles generally forbid a tree level cubic $\phi^3$ term in the potential.  However, finite temperature corrections to the potential generally scale as $V_{\rm 1-loop} \propto T^4 \int_0^\infty dx \, x^2 \ln(1 \mp e^{- \sqrt{ x^2 + y}})$, where $y = m^2 \slash T^2$.  In the high temperature limit, the expansion of the logarithm generates terms $\propto T |\phi|^3$ (assuming the mass is proportional to the scalar VEV).  Furthermore, because the coefficient is generically negative, the potential grows slower than quadratically when this term dominates. 

\paragraph{}The Standard Model provides an excellent playground to explore these states.  Because the Higgs field is responsible for mass generation, the finite temperature corrections have the necessary form.  Furthermore, because the running quartic term is small, there is a comparatively large regime in which the cubic term is relevant.  In this paper, we will explore the properties of these ``Higgs balls".  We emphasize that these non-topological solitons are made of Higgs quanta; this is different from other models in which non-topological solitons form from other scalar quanta whose interactions are mediated by Higgs bosons~\cite{Endo:2022uhe,Bishara:2017otb}.

\paragraph{}However, one complication is that the Standard Model Higgs carries gauged, not global, charge.  When a field carries gauged (as opposed to global) symmetry, Q-balls may still exist~\cite{Rosen:1968mfz,Lee:1988ag,Lee:1991bn}, although only for sufficiently small charges~\cite{Lee:1988ag,Gulamov:2015fya}.  (See alose the review Ref.~\cite{Nugaev:2019vru}.)  As we discuss, this has implications for the existence of Q-balls, both in the Standard Model and in extensions.  We also focus on Higgs balls because they illustrate many non-trivial aspects that can arise in thermal balls.  In addition to being gauged (not global) Q-balls, they are also not absolutely stable.

\paragraph{}Like global Q-balls, gauged Q-balls may be long-lived but not absolutely stable.  As mentioned, our Higgs balls will be of this type, because the Higgs quanta can decay (for example, through the Yukawa interactions).   Since the Higgs field is the only scalar field that carries $\mathrm{SU}(2)$ and $\mathrm{U}(1)$ charge, we note that we can consider the Higgs balls as carrying a global ``Higgs charge''.  When the rate of Higgs-particle-number-changing processes inside the soliton is smaller than the Hubble parameter, Higgs charge is effectively conserved. Therefore, we show the Higgs balls are long lived for sufficiently high temperatures.  This can be generalized to other gauged thermal balls.

\paragraph{}This paper is organized as follows: First, in section \ref{sec:thermal_balls} we discuss how thermal effects can produce Q-balls at finite temperature in models which do not have Q-balls at zero temperature.  We then turn to Higgs balls as a specific illustration of these ``thermal balls".   In section \ref{sec:SM_Hballs}, we first study Higgs balls in a Standard Model-like scenario in which the $\mathrm{U}(1) \times \mathrm{SU}(2)$ symmetry is a global symmetry.  We see that the finite temperature corrections allow for the existence of (global) Q-balls.

\paragraph{}In section \ref{sec:gauging}, we then include gauge interactions by generalizing results for $\mathrm{U}(1)$ gauged Q-balls to $\mathrm{SU}(2) \times \mathrm{U}(1)$.  We see that even in the static charge limit, the gauge interactions are sufficient to destabilize Higgs balls in the Standard Model; they have higher energy than an equivalent number of free Higgs quanta.

\paragraph{}However, it is not necessarily the case that the gauge interactions always destabilize the Q-balls, and in section \ref{sec:BSM_Hballs}, we consider an extension of the Standard Model in which non-topological solitons remain stable against decay into separate Higgs quanta.

\paragraph{}Finally, although the Higgs balls are stable against decay into individual Higgs quanta, the Standard Model Yukawa couplings enable them to decay into fermions.  This decay rate is suppressed since it can only occur at the surface, and we make brief remarks on this in section \ref{sec:decay}. 

\paragraph{}Finally we conclude, emphasizing both the general features that enable the existence of ``thermal balls" in various models, as well as the specific features of Higgs balls.

\section{Thermal Balls}
\label{sec:thermal_balls}

\paragraph{}In this section, we discuss how thermal effects can generally create non-topological solitons in models which have no Q-balls at zero temperature.  
\paragraph{}We consider a scalar field $\phi$ with vacuum expectation value (VEV) $v$, which at zero temperature is determined by minimized a potential $V(v)$.  The one-loop finite temperature corrections to this potential can be expressed as~\cite{Arnold:1992rz, DelleRose:2015bpo}
\begin{align}
V_{1-\mathrm{loop}}(v,T) 
&= 
	\sum_{\mathrm{bosons}} \dfrac{n_iT^4}{2 \pi^2} J_B \left( \dfrac{m_i^2}{T^2} \right) 
	+ \sum_{\mathrm{fermions}} \dfrac{n_i T^4}{2 \pi^2} J_F \left( \dfrac{ m_i^2}{T^2} \right),
\end{align}
where 
\begin{align}
	J_B(y) &= \int_0^\infty dx \, x^2 \ln \left( 1 - e^{- \sqrt{x^2+y} }\right) , \nonumber \\
	J_F(y) &= \int_0^\infty dx \, x^2 \ln \left( 1 + e^{- \sqrt{x^2+y} }\right) .
\end{align}

\paragraph{}Although we use the full finite temperature corrections, the high temperature limit is helpful in understanding this phenomenon.  In this limit, the functions above become
\begin{align}
J_B(x) &= - \dfrac{\pi^4}{45} + \dfrac{\pi^2}{12} x - \dfrac{\pi}{6} x^{3 \slash 2}, \nonumber \\
J_F(x) &= \dfrac{7 \pi^4}{360} - \dfrac{\pi^2}{24} x,
\end{align}
with the next order corrections of order $x^2 \ln(x)$.  If the bosons have a mass $\sim g v$, the bosonic loops will produce a term of the form $- g^3 T |v|^3$, which is effectively an attractive interaction.

\paragraph{}As has been noted~\cite{Buchmuller:1993bq}, second order loop corrections also contribute to this cubic term, due to the regularization of infrared divergences.  These can be calculated through the ring (or daisy) diagrams, which we include in our calculations below.  We also note that although the conceptual explanation above makes use of the high temperature expansion, in the results presented below we numerically integrate the $J_B$ and $J_F $ functions in the regime in which the high temperature expansion is not valid.

\paragraph{}The importance of these cubic terms in spontaneous symmetry beaking has long been noted; when they are sufficiently large the phase transition is first order~\cite{Buchmuller:1993bq}.  In this work, we demonstrate their importance to the existence of solitons.

\paragraph{}The bosonic contributions can come from either scalar or vector fields.  Scalar fields can have a mass term generated by interactions with the $\phi$ field.  We assume that $\phi$ carried a global charge, in which case non-topological solitons (Q-balls) exist if $V(v,T) \slash v2$ is minimized at non-zero $v$.  We see immediately that the existence of a $-A g^3 T |v|^3$ ensures that this is satisfied.  We note that this can occur even in a model in which $\phi$ is the only field, provided that at sufficiently large VEVs its mass is given by its self-interaction times its VEV.  As we will see below, in the Standard Model the small Higgs self-coupling suppressed this contribution, but this will be explored further in Section~\ref{sec:BSM_Hballs}, when we consider BSM (beyond the Standard Model) scenarios.

\paragraph{}This situation is more complicated when vector bosons are used.  Vector bosons generally acquire masses of the form $gv$ through spontaneous symmetry breaking, in which case the scalar field $\phi$ carries gauge charge.  The gauge fields mediate self-interactions, which can increase the energy of the Q-ball.  If the gauge group in question is $\mathrm{U}(1)$, the energy per unit charge of the gauged Q-ball (in the thin wall limit) is given by~\cite{Heeck:2021zvk}
\begin{align}
\omega(R) &= \omega_0(R) g v_0 R  \coth(g v_0 R )
\label{eq:U1_qballs}
\end{align}
where $\omega_0$ is the energy per unit charge of a global Q-ball (i.e., calculated ignoring gauge self-interactions of the scalar field), and $v_0$ is the VEV inside this Q-ball.  Provided that $\omega(R) < m_\phi$, the mass of a scalar quanta outside the soliton, non-topological soliton solutions exist.  For sufficiently small $g v_0 R$, the additional energy from gauge-self interactions is small and non-topological solitons will exist.  

\paragraph{}However, $\omega_0$ (and $v_0$) will generally depend on $g$, and therefore the existence of gauged thermal Q-balls has to be considered in each individual model.  Since the only scalar field confirmed to exist is the Higgs field, we are particularly interested in it, which requires us to generalize the above results.

The second condition for a Q-ball is traditionally the presence of a conservation law for particle number, such that the quanta cannot lose energy by decaying into lighter particles. In the early Universe, we argue that this condition is too strict and one merely requires that the decay rate of the quanta be smaller than the expansion rate of the Universe.

\paragraph{}We will call the non-topological solitons produced by thermal corrections, and therefore which do not exist at zero temperature, thermal balls.  We now turn our attention to determining whether such thermal balls exist in the Standard Model Higgs sector.

\section{Higgs Balls in the Global Standard Model}
\label{sec:SM_Hballs}

\paragraph{}As the Standard Model of particle physics has a single scalar field, the Higgs field, it is natural to first ask whether these exist within the Standard Model.  To answer this question, we first study the non-topological solitons that would exist if the Standard Model $\mathrm{SU}(2) \times \mathrm{U}(1)$ symmetry were not gauged.  More specifically, we ignore repulsive interactions mediated by the gauge bosons, which allows us to use global soliton equations.  (We do keep all Standard Model interactions when calculating corrections to the Higgs potential, including those involving the gauge bosons.) We do this for two reasons: first, to show the qualitative features that determine the existence of thermal balls, and secondly, as we demonstrate in section \ref{sec:gauging}, the energy of gauged Higgs balls can be found using the energy of this global model.  (This is expected from the $\mathrm{U}(1)$ results in Ref.~\cite{Heeck:2021zvk}; see equation \eqref{eq:U1_qballs}.)  

\paragraph{}In this section, we consider the Standard Model Higgs, whose running quartic coupling becomes small (and even negative) at large field values before increasing again, and therefore there is a regime in which the potential grows slower than quadratically.  If the Higgs field carried only global charge, then $V(h)\slash h^2$ would be minimized at non-zero $h$, indicating the existence of thin wall Q-balls~\cite{Coleman:1985ki}. However, such Q-balls would typically have $V(h) < 0$ in their interiors, and they would induce a phase transition to the true vacuum at large VEVs via the process of solitosynthesis~\cite{Kusenko:1997hj,Postma:2001ea,Metaxas:2000qf,Pearce:2012jp}).  We do not consider these states.

\paragraph{}The measured top quark pole mass is $173.1 \pm 0.9 \, \mathrm{GeV}$, and for the central value, the Higgs potential becomes unstable around $10^{10} \, \mathrm{GeV}$.  However, the scale of the instability depends strongly on the top quark mass.  In order to better illuminate our solitons, we will impose that the Higgs potential is stable up to $10^{17} \; \mathrm{GeV}$, which corresponds to a pole top quark mass of $170.5 \, \mathrm{GeV}$, which is within $3 \sigma$ of current observations~\cite{ParticleDataGroup:2020ssz}.

\paragraph{}To generate our solitons, we note that in the early universe, finite temperature corrections modify the Higgs potential~\cite{Quiros:1994dr}.  In the work the follows, we have included one-loop corrections to the Standard Model potential and one-loop and ring (``daisy'') finite temperature corrections.  The renormalization group equations governing the running coupling constants have been evaluated to two loop order.  For the finite temperature corrections, we include contributions from the gauge bosons as well as the Higgs and the top quark, but other fermions are neglected due to their small Yukawa masses.    Details on how we calculated the finite temperature Standard Model potential and evaluated the running coupling constants can be found in Appendix \ref{ap:SM_potential}, and details of how we matched the parameters to observables are in Appendix \ref{ap:matching}.

\begin{figure}
    \centering
    \includegraphics[scale=0.6]{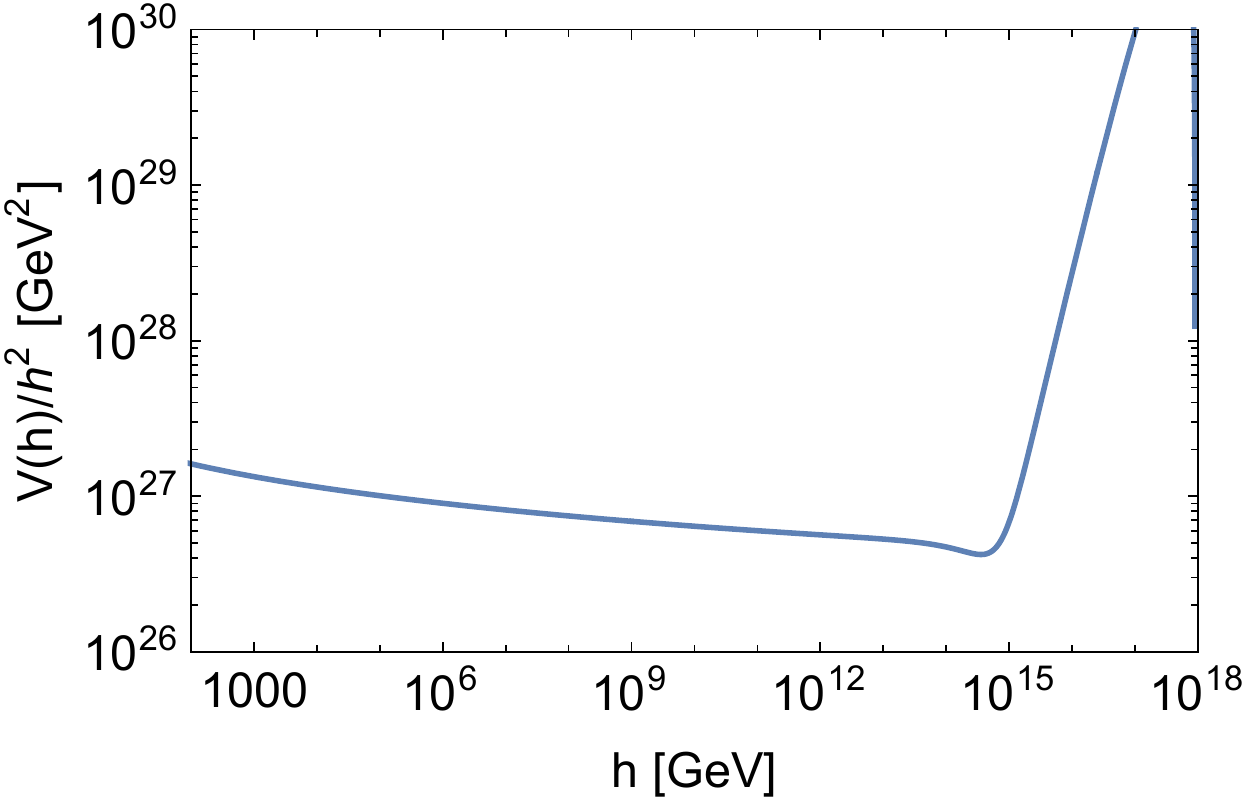}
        \includegraphics[scale=0.56]{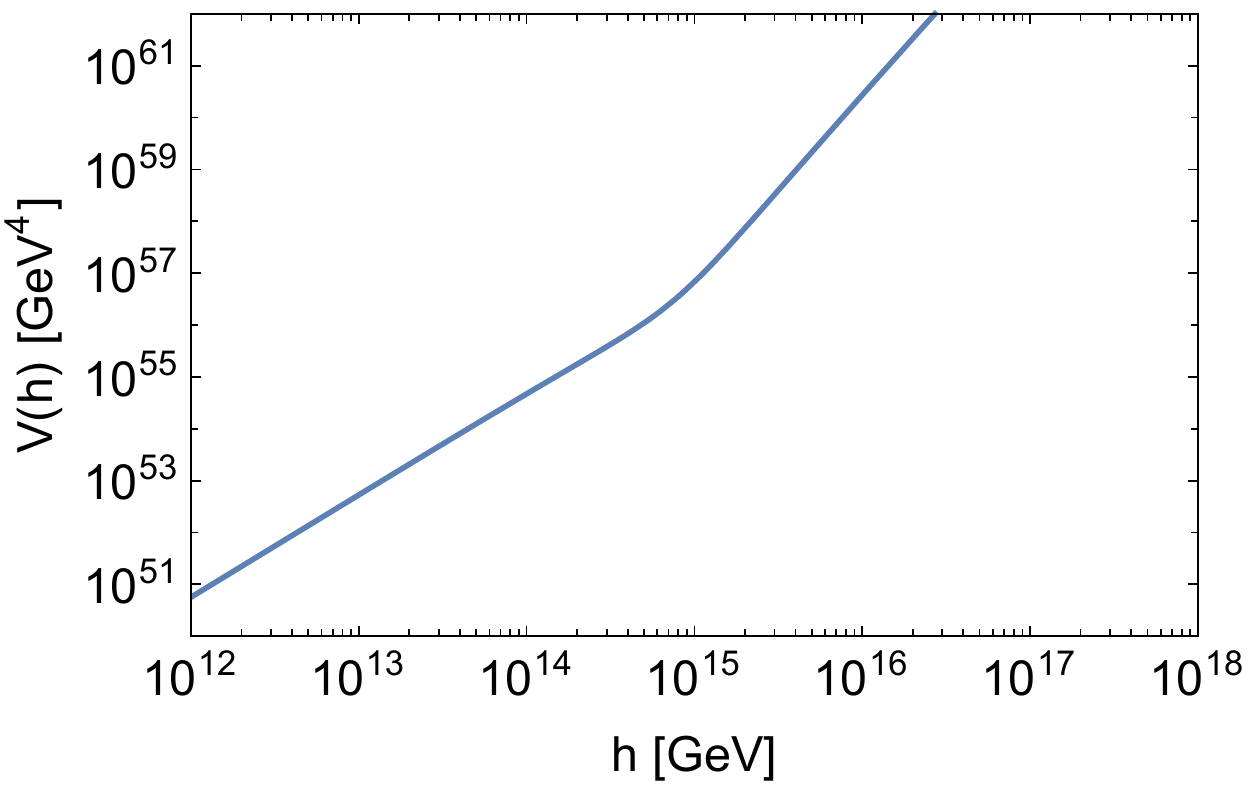}
    \caption{Left: $V(h) \slash h^2$ as a function of Higgs VEV $h$, at a fixed temperature of $T = 10^{14} \, \mathrm{GeV}$.  Because this is minimized at nonzero VEV, thin wall Q-ball solutions exist.  Right: $V(h)$ as a function of Higgs VEV, at a fixed temperature of $T = 10^{14} \, \mathrm{GeV}$.  The Higgs potential remains positive throughout the regime of interest, avoiding solitosynthesis.}
    \label{fig:global_existence}
\end{figure}

\paragraph{}The left side of Fig.~\ref{fig:global_existence} shows the $V(h) \slash h^2$ at temperature $T = 10^{14} \, \mathrm{GeV}$, and is clearly minimized at nonzero field values $h \sim 10^{15} \, \mathrm{GeV}$.  To demonstrate that this is not a solitosynthesis scenario, we show the potential $V(h)$ as a function of the Higgs VEV on the right; as expected, it is positive up to the stabilization scale of $10^{17} \, \mathrm{GeV}$.  

\begin{figure}
    \centering
    \includegraphics[scale=0.6]{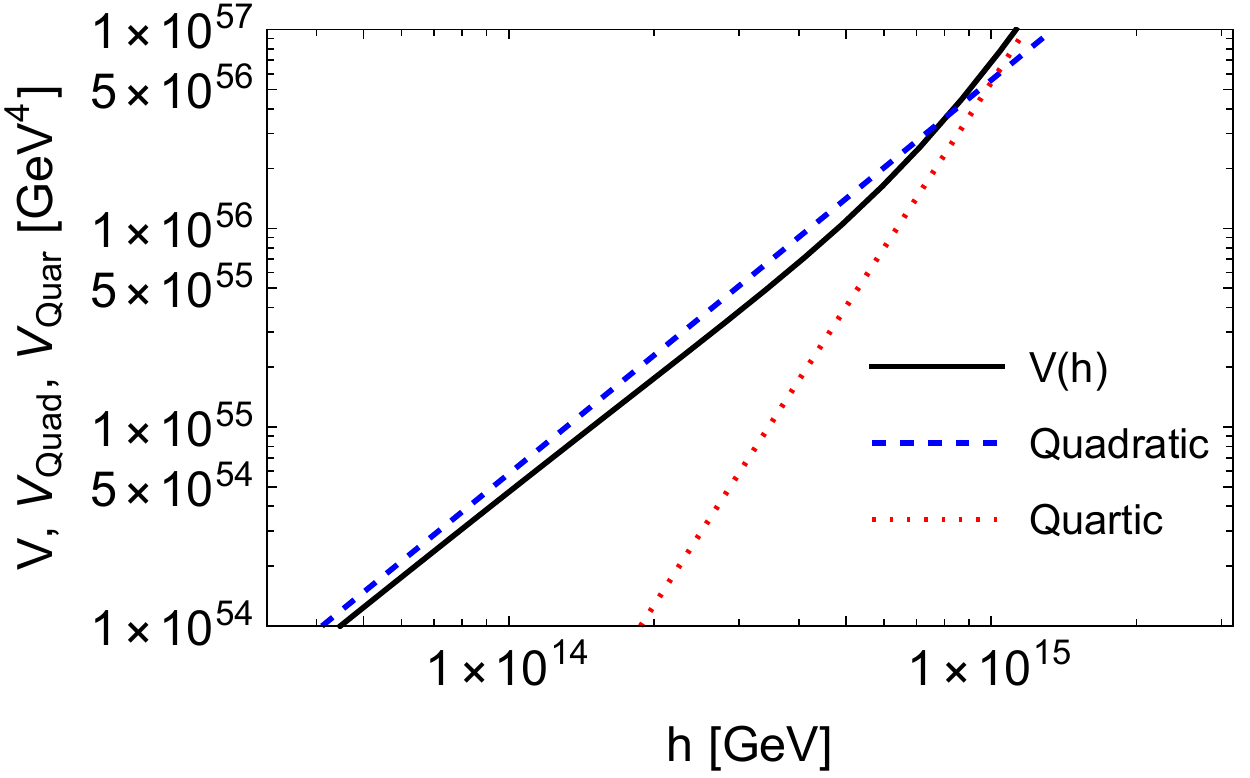}
    \caption{The potential $V(h)$ (black) compared to the quadratic (blue, dashed) and quartic (red, dotted) terms at $T= 10^{14} \, \mathrm{GeV}$.  Before the quartic terms dominate around $h = 10^{15} \, \mathrm{GeV}$, the potential is below the quadratic contribution due to finite temperature corrections.}
    \label{fig:global_cubic}
\end{figure}

\paragraph{}Fig.~\ref{fig:global_cubic} illustrates concretely how finite temperature contributions lead to the Q-ball minimum.  This figure compares the potential to the temperature-dependent quadratic and quartic terms.  At low VEVs, we see that the temperature-dependent quadratic term dominates.  However, as the VEV increases, the potential begins to drop below the quadratic terms, indicating the existence of thin wall Q-balls.  This continues until $h \sim 10^{15} \, \mathrm{GeV}$, when the quartic term dominates.

\begin{figure}
    \centering
    \includegraphics[scale=0.57]{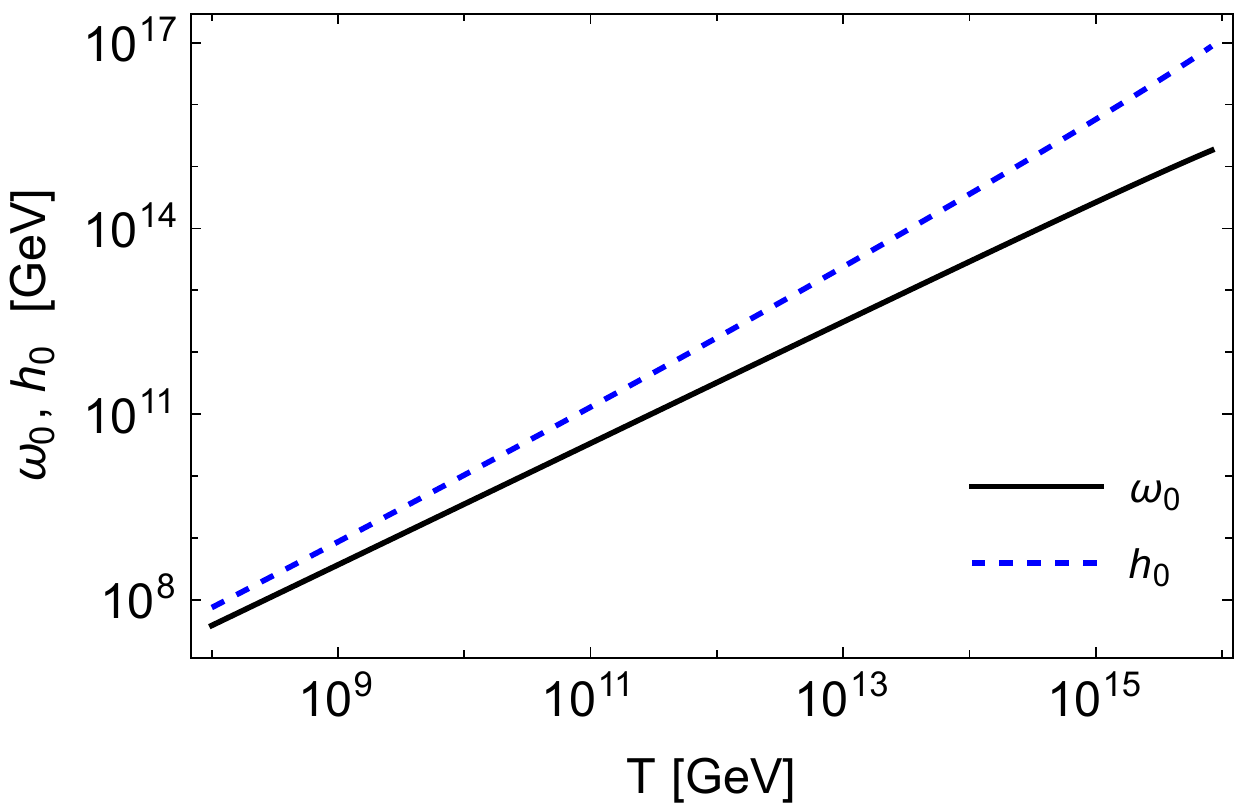}
    \caption{The energy per unit charge $\omega_0$ of the thin-wall Higgs ball (black) and the Higgs VEV $h_0$ inside the Higgs ball (blue, dashed) as a function of temperature.}
    \label{fig:global_omega_vev}
\end{figure}

\paragraph{}Before discussing gauge effects, it is helpful to study the properties of these Q-balls, including their stability.  Despite being large classical states, they behave like radiation.   The left plot of Fig.~\ref{fig:global_omega_vev} shows the energy per unit charge of the Higgs ball, as a function of temperature, and we see that it scales linearly.  This plot also shows the Higgs VEV in the interior of the thin wall Q-ball, determined by minimized $V(h,T) \slash h^2$.  This is also approximately linear in temperature, although the linearity begins to break down at larger scales.

\begin{figure}
    \centering
    \includegraphics[scale=0.57]{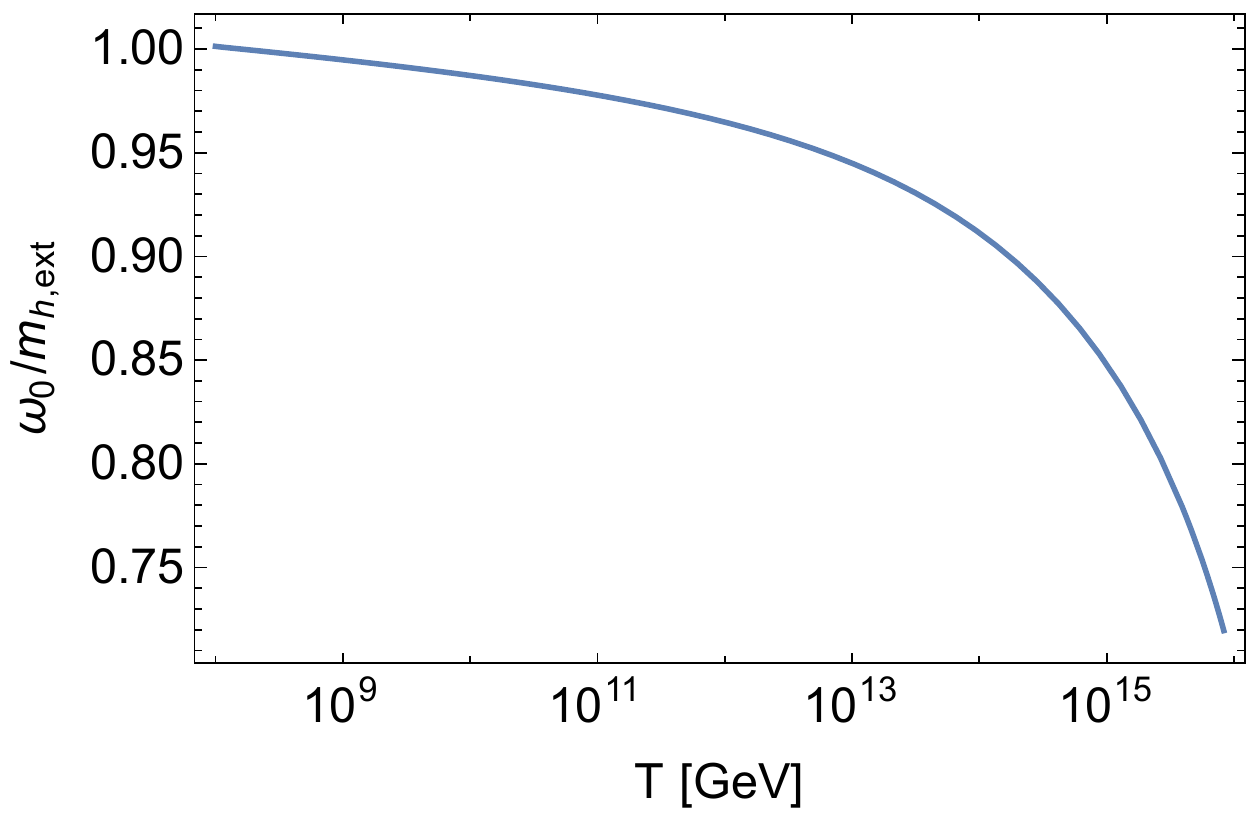}
    \includegraphics[scale=0.57]{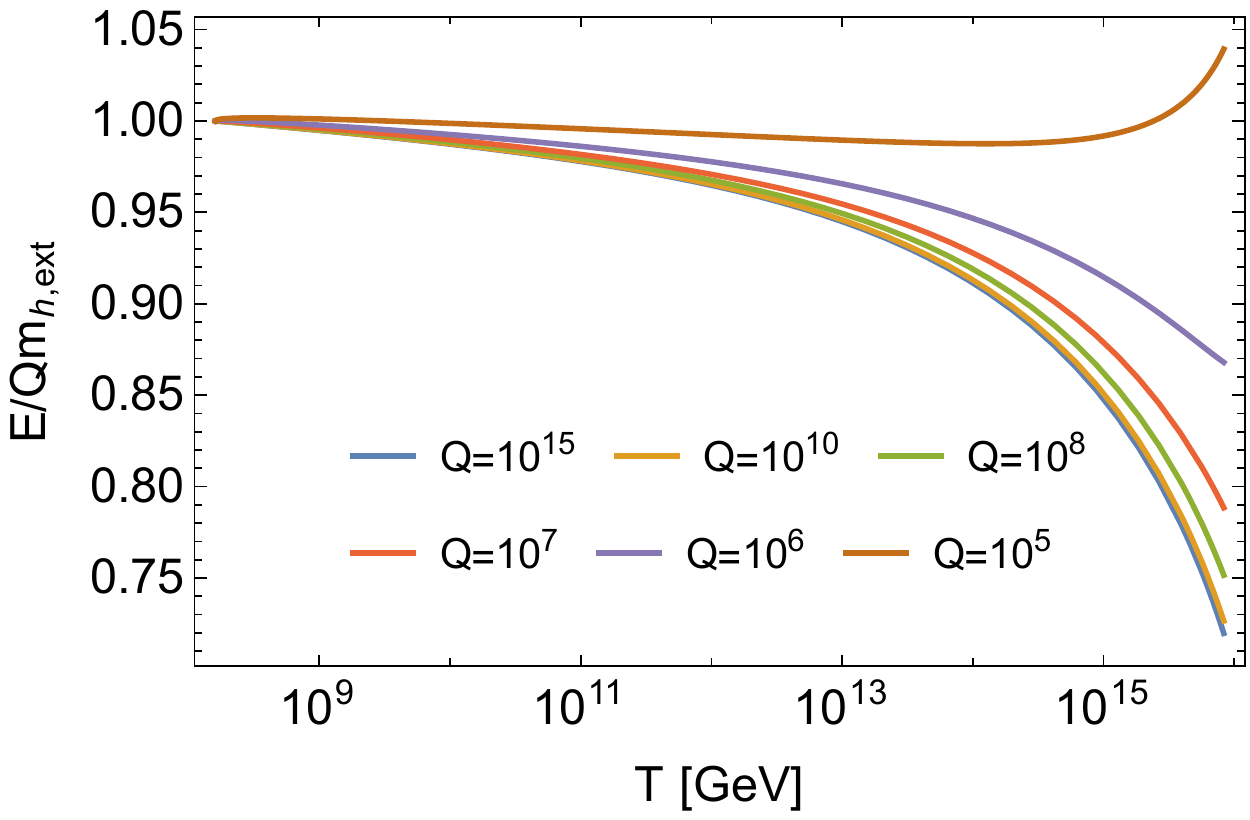}
    \caption{Left: The ratio of the energy per unit charge of the Higgs ball to the mass of a Higgs quanta outside of the Higgs ball, as a function of temperature.  The Higgs ball is stable when this is less than 1.  Right: The total energy of the Higgs ball, including surface effects, as a function of temperature for a variety of charges. (Color online.)}
    \label{fig:global_stability}
\end{figure}

\paragraph{}Finally, we discuss the stability of the Higgs balls.  Fig.~\ref{fig:global_cubic} shows that the $V(h,T) \slash h^2$ is only slightly smaller than the quadratic term, which is likewise dominated by finite temperature corrections.  Frequently this would be sufficient to show that the Higgs ball has lower energy the equivalent number of free quanta, but we must be careful when evaluating the mass of the free Higgs quanta.  Inside the Higgs ball, we have used the maximum of the temperature or Higgs VEV as our renormalization scale for our running couplings, but in the plasma outside the Higgs ball the temperature should be used.  Therefore, the mass of a Higgs quanta outside the Q-ball is not necessarily the same as the one determined from the quadratic contribution to the potential inside the Q-ball.  We compare the energy per unit charge $\omega_0$ to the mass of a Higgs quanta outside the Q-ball in the left plot of Fig.~\ref{fig:global_stability}, and see that the Q-ball is stable against decay into free Higgs  quanta for $T \gtrsim 10^{10} \, \mathrm{GeV}$.  (As we will discuss below, Yukawa couplings still allow these to decay into fermions, although they are long-lived.)

\paragraph{}This ensures the stability of thin wall Q-balls, in which the surface tension contribution is neglected.  Given the small difference between $\omega_0$ and $m_{h,{\rm ext}}$, we should be concerned that the small contribution to the energy from the surface tension may be sufficient to destabilize them.  (By this, we mean that the energy per unit charge is greater than the mass of a Higgs quanta outside the Q-ball.)  For a Q-ball of charge $Q$, the total energy including the surface contribution is
\begin{align}
E &= Q \omega_0 + \dfrac{4\pi}{3} h_0^2 R \cdot \sqrt{ m_{h,{\rm ext}}^2 - \omega_0^2},
\end{align}
and we calculate the radius using the thin wall approximation, since we are in the regime in which the surface tension energy is small compared to $Q \omega_0$.  This is plotted for a variety of charges in the right plot of Fig.~\ref{fig:global_stability}, showing that for charges $Q \gtrsim 10^6$, surface tension is not sufficient to destabilize the Q-ball.

\paragraph{}Therefore, in the (unphysical) scenario in which the Standard Model $\mathrm{SU}(2) \times \mathrm{U}(1)$ is ungauged, we see that thermal effects produce non-topological solitons with charge greater than $10^5$ in the temperature range $10^{10} \, \mathrm{GeV}$ and $10^{14} \, \mathrm{GeV}$.

\section{Gauge Effects: Higgs Balls in the Standard Model}
\label{sec:gauging}

\paragraph{}The Standard Model Higgs carries gauged, not global charge.  Q-balls carrying gauged $\mathrm{U}(1)$ charge have been studied in Ref.~\cite{Heeck:2021zvk}, which showed that in the thin wall limit the energy per unit charge of the gauged Q-ball $\omega$ is related to that of a global Q-ball, $\omega_0$.  In Appendix \ref{ap:gauging_qballs}, we generalize these first to a scalar field carrying gauged $\mathrm{SU}(2)$ charge, and then to the Higgs field, which carries both $\mathrm{SU}(2)$ and $\mathrm{U}(1)$ gauged charge.

\paragraph{}We note that as in Ref.~\cite{Heeck:2021zvk}, we make the static charge approximation.  This approximation is perhaps more problematic for gauged charge, as the assumption that only the zero component of the gauge fields is nonzero removes $\mathrm{SU}(2)$ self-interactions between the gauge fields in the Q-ball.  The breakdown of the static charge approximation is connected to the confining aspect of $\mathrm{SU}(N)$ gauge interactions.  In this work, we will ensure that the radii of our Q-balls is less than the $\mathrm{SU}(2)$ confinement scale, as calculated outside the Q-ball where the symmetry is unbroken.  We discuss this further in Appendix \ref{ap:gauging_qballs}.

\paragraph{}As shown in the appendix, the energy per unit charge of the Higgs ball is related to that of the ungauged Higgs ball by
\begin{align}
\omega =\frac{1}{2} R h_0 \omega_0 \sqrt{g_W^2+g_Y^2} \coth \left(\frac{1}{2} R h_0 \sqrt{g_W^2+g_Y^2}\right).
\label{eq:gauged_omega}
\end{align}

A Q-ball made of the Higgs field will carry both $\mathrm{U}(1)$ and $\mathrm{SU}(2)$ charge; the amount of $\mathrm{SU}(2)$ charge is described by the weak isospin.  A quanta of the Higgs field has hyperchange $1 \slash 2$ and weak isospin $-1 \slash 2$, so the charges of the Higgs ball satisfy $Q_Y = - Q_W = - \dfrac{i}{2} \int d^3x \, \left(  \Phi^\dagger D_0 \Phi - h.c. \right)$, as it must for it to be electrically neutral.  As shown in the appendix, the charge is related to the large radii behavior of the gauge fields.  As shown in the appendix,
\begin{align}
Q_Y = - Q_W &= \frac{8 \pi  R \omega_0 \left(R h_0 \sqrt{g_W^2+g_Y^2} \coth \left(\frac{1}{2} R h_0 \sqrt{g_W^2+g_Y^2}\right)-2\right)}{g_W^2+g_Y^2}.
\label{eq:gauged_charge}
\end{align}
We note that the Appendix includes explicit expressions for the gauge fields are derived in the static charge limit, thin wall case.

\begin{figure}
    \centering
    \includegraphics[scale=0.55]{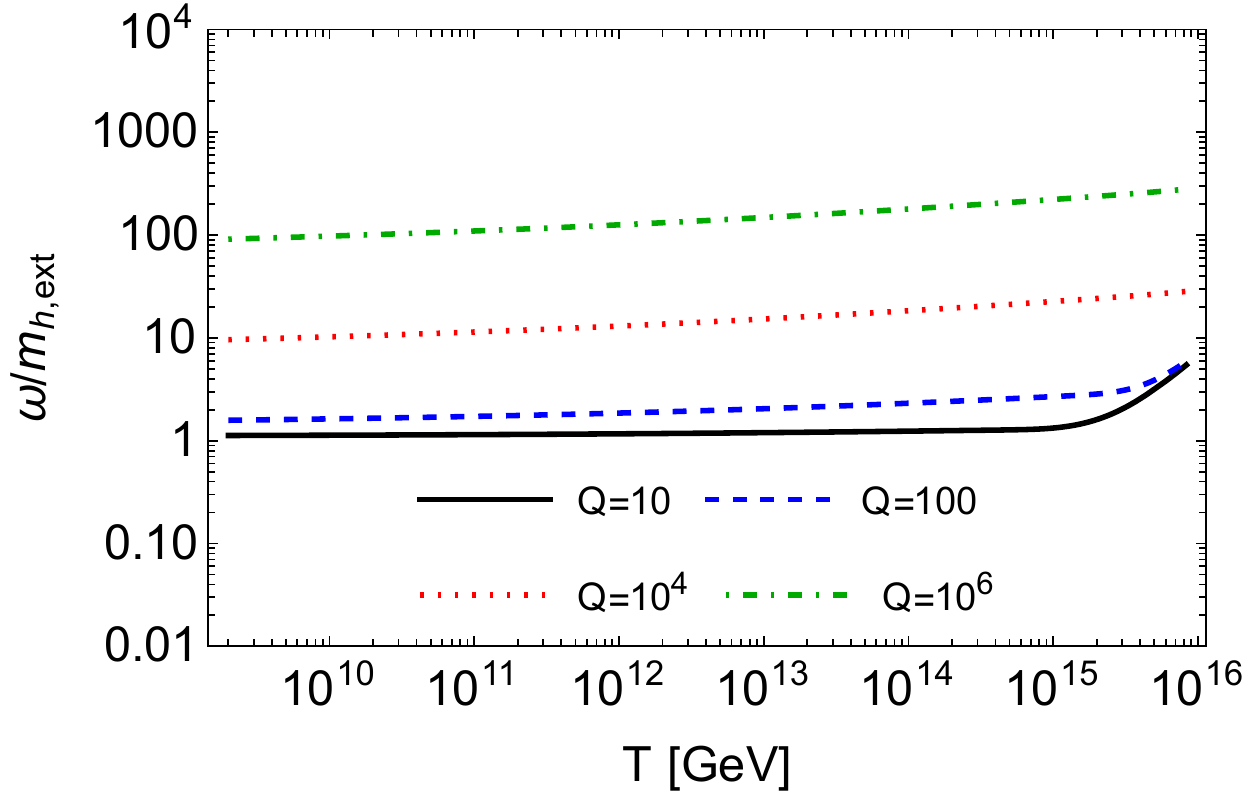}
    \caption{Ratio of energy per unit charge to Higgs mass outside the Q-ball, as a function of temperature, for a range of charges.  Because this ratio is larger than one, if such Higgs balls were produced they would immediately decay into individual quanta.}
    \label{fig:gauged_unstable}
\end{figure}

\paragraph{}For a chosen charge $Q=|Q_Y|=|Q_W|$, equation \eqref{eq:gauged_charge} can be numerically solved to find the radius of the gauged Q-ball, which can then be substituted into equation \eqref{eq:gauged_omega} to find the energy per unit charge, including the potential energy from gauge interactions between the Higgses.  To determine stability, this should be compared to the mass of a Higgs quanta in the plasma outside the Higgs ball.  This is shown in Fig.~\ref{fig:gauged_unstable}, which shows that the ratio is generally larger than one.  This means the Q-ball could lower its energy by becoming separate free Higgs quanta.  In fact, such states would not be produced, as the repulsion force mediated by the gauge bosons would push the Higgs quanta apart, to the lower energy state.

We could solve for the radius at which the energy per unit charge for the gauged Q-ball is equal to the external Higgs mass, and then use equation \eqref{eq:gauged_charge} to determine the charge of this state.  Across the temperature range of interest it is $\mathcal{O}(0.1)$, well outside the thin wall regime of validity.  Therefore, we conclude that Higgs balls do not exist in the Standard Model, as the contribution to the energy from gauge interactions is sufficient to destabilize them.

\section{Higgs Balls Beyond the Standard Model}
\label{sec:BSM_Hballs}

\paragraph{}As the above makes clear, non-topological solitons cannot be constructed with the Standard Model Higgs because the increased energy from gauge interactions makes them unstable into decay into individual Higgs quanta.  However, there is no fundamental reason why such states could not exist, and since the behavior of the Higgs potential at high scales is unknown, the existence of such states is an open question.  In this section, we explore Higgs balls in an extension of the Standard Model.  

\paragraph{}Higgs balls are unstable due to the increased energy from gauge self-interactions; these are decreased if the gauge couplings $g_Y$ and $g_W$ decrease.  Although this can be accomplished by changing their running (for example, through additional fermionic contributions with a negligible coupling to the Higgs, which would provide a negative contribution to the $beta$ functions of the gauge couplings while leaving the Higgs quartic unaffected), this is insufficient to stabilize the Higgs balls.  The finite temperature corrections which produce the minimum in $V(h) \slash h^2$ arise from the bosonic $J_B$ functions and scale as powers of the bosonic masses.  The contributions of the gauge bosons disappear as the gauge couplings decrease, leaving only the much suppressed Higgs and ghost contributions.

\paragraph{}Therefore, in addition to modifying the gauge couplings at large scales, we additionally include a singlet field $S$.  The potential is
\begin{align}
V(H,S) = - \mu^2 H^\dagger H + \lambda_H (H^\dagger H)^2 + \dfrac{m_S^2}{2} S^2 + \lambda_S S^4 + \lambda_{HS} H^\dagger H S^2,
\end{align}
where we have imposed a $\mathbb{Z}_2$ symmetry on the scalar.  We assume the mass of the scalar singlet $m_S$ is above the LHC scale, but when the Higgs acquires a large VEV $h$ the singlet's mass is given by
\begin{align}
m_{S,\mathrm{eff}} &= \sqrt{ m_S^2 + \lambda_{HS} h^2} \approx \sqrt{\lambda_{HS}} h.
\end{align}
This mass is independent of the gauge couplings, which consequently decouples the depth of the $V(h) \slash h^2$ minimum from the gauge coupling.  We note that the scalar field $S$ does not acquire a vacuum expectation value and will not mix with the Standard Model Higgs boson.  Furthermore, in our benchmark scenario we will take $\lambda_{HS} = 0.9$, forbidding Higgs decays into these scalars even at large VEVs.

\paragraph{}Now that we have decoupled the depth of the H-ball minimum from the gauge couplings, we modify the running gauge couplings.  We assume that at some scale above the LHC scale but below the temperatures of interest here, new states cause the Standard Model gauge couplings to run to significantly smaller values.  As noted above, this can be implemented through the introduction of additional fermions; if these carry gauge charge, they contribute to the $g_Y$ and $g_W$ gauge functions.  We note that at least at the one loop level, it is impossible for the running couplings to become negative, as each beta function is proportional to the cube of the coupling.  Consequently, it becomes arbitrarily small.  

\paragraph{}We note that the fermions do not necessarily couple to the Standard Model Higgs, although in the absence of a new gauge interaction such couplings are generically expeccted.  We assume that either a new gauge interaction forbids such couplings or the couplings are exceedingly small, so that the effect of the new fermions on the Higgs potential is negligible.  

\paragraph{}Since the only impact of the fermions is to decrease the running gauge couplings from their relatively large Standard Model values, we do not present a complete discussion of the fermionic sector, but rather parameterize our results by the decrease in the gauge couplings.  We will focus primarily on a benchmark in which the gauge couplings at large VEVs are $1 \slash 100$th their running Standard Model values, although we will also briefly discuss the case in which they are $1 \slash 10$th their running Standard Model values.  Finally, to preserve the running of the Standard Model Higgs quartic coupling to small values over many orders of magnitude, we also introduced a fermion which at the same scale as the singlet, with a Yukawa coupling related $\lambda_{HS}$.  Full details of the modifications made in this scenario are in Appendix \ref{ap:BSM_Changes}.

\paragraph{}Fig.~\ref{fig:BSM_global_cubic} is the analog of Fig.~\ref{fig:global_cubic} in our beyond-the-Standard-Model (BSM) scenario.  By comparison, we see that the new scalar field causes the full potential to be beneath the quartic contribution, indicating the existence of Q-balls.  We also see that the potential is positive throughout this region, and thus this scenario is protected from solitosynthesis.  Although this plot may appear similar to the Standard Model plot in Fig.~\ref{fig:global_cubic}, we emphasize that the source of the relevant finite temperature terms is different.  In the pure Standard Model scenario, the gauge bosons were responsible for the dip.  Here, their contribution is negligible because the gauge couplings are significantly smaller.  Instead, the dip is caused by the scalar field.  However, the effective coupling $\lambda_{H S}$ is of order $\mathcal{O}(0.1)$, the same order as the (original) Standard Model gauge couplings. For this reason, the dip produced by the singlet in our BSM scenario is comparable to the dip produced in the pure Standard Model scenario produced by the gauge bosons.  (Additionally, the fact that we adjust the top pole mass to the maximum value consistent with the potential being stable to scales of $10^{17} \, \mathrm{GeV}$ also contributes to the similar appearance of the potentials in the two scenarios.)

\begin{figure}
    \centering
    \includegraphics[scale=0.6]{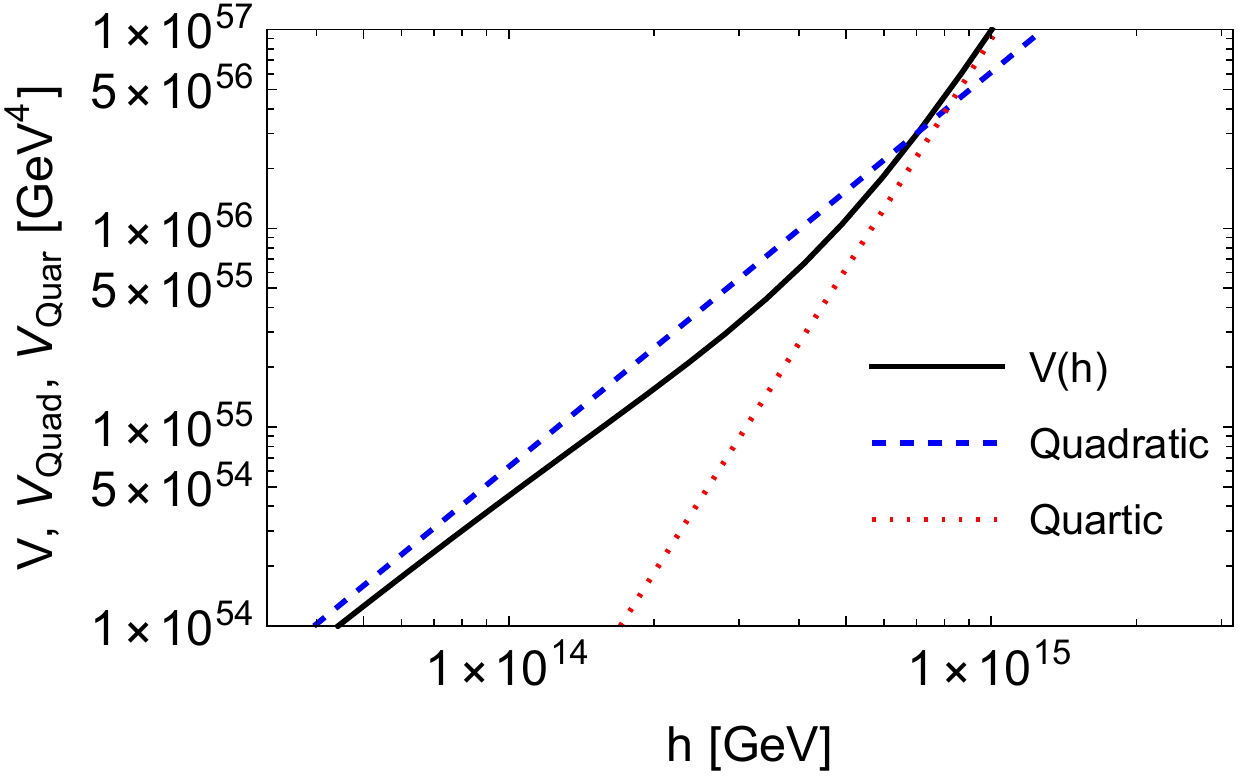}
    \caption{Left: The potential $V(h)$ (black) compared to the quadratic (blue, dashed) and quartic (red, dotted) terms at $T= 10^{14} \, \mathrm{GeV}$, for our beyond-the-Standard-Model scenario.  Before the quartic terms dominate around $h = 10^{15} \, \mathrm{GeV}$, the potential is below the quadratic contribution due to finite temperature corrections.}
    \label{fig:BSM_global_cubic}
\end{figure}

\begin{figure}
    \centering
    \includegraphics[scale=0.57]{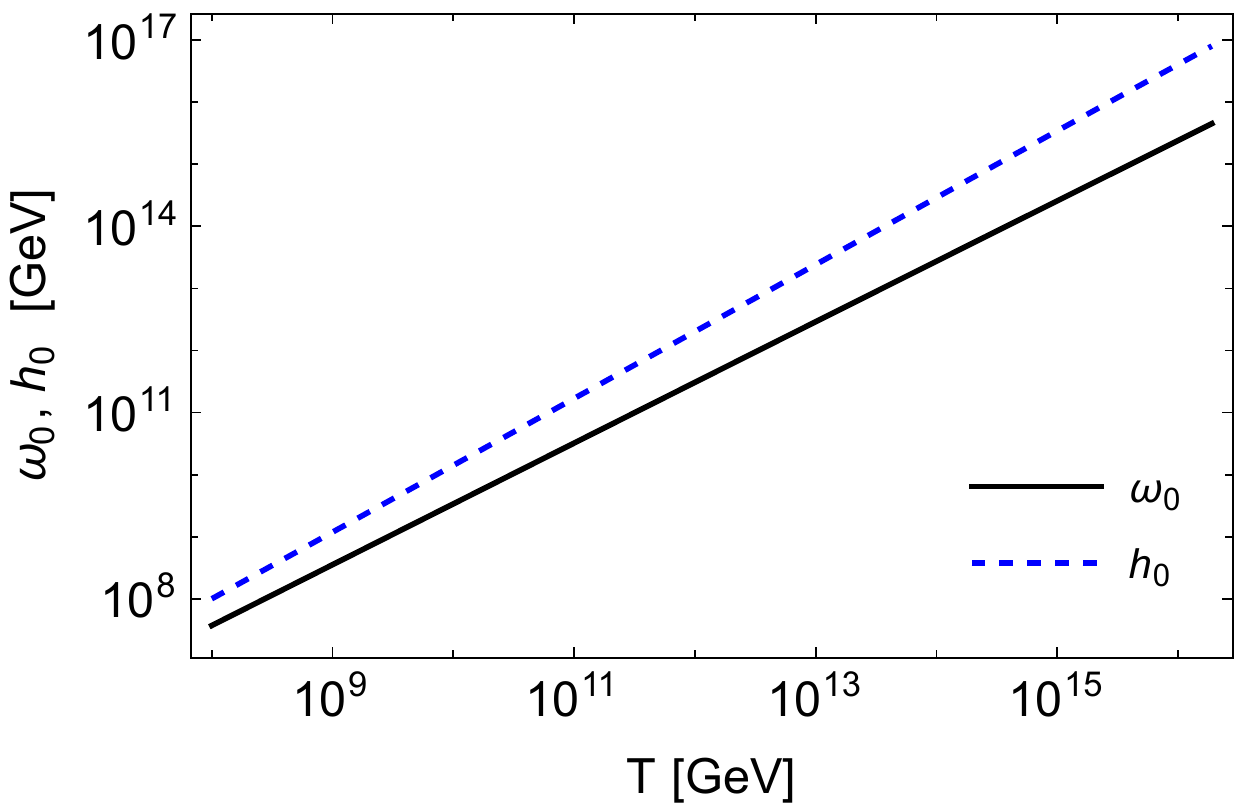}
    \includegraphics[scale=0.57]{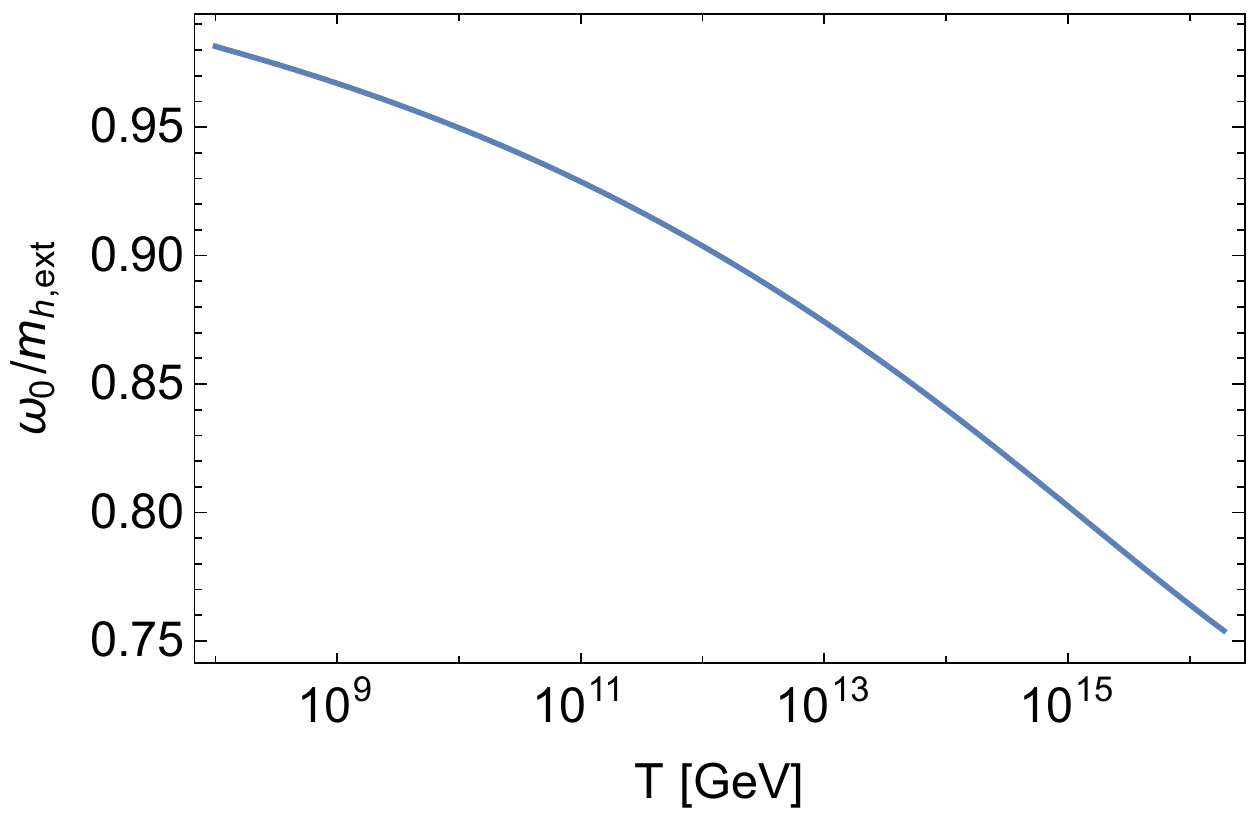}
    \caption{Left: The energy per unit charge $\omega_0$ of the thin-wall global Higgs ball (black) and the Higgs VEV $h_0$ inside the global Higgs ball (blue, dashed), as a function of temperature, for our BSM scenario.  Right: The ratio of the energy per unit charge of the global Higgs ball to the mass of a Higgs quanta outside of the Higgs ball, as a function of temperature.  The global Higgs ball is stable when this is less than 1.}
    \label{fig:BSM_global_omega_vev}
\end{figure}

\paragraph{}As discussed above, gauged Q-balls are closely related to their global counterparts, and therefore as above, we first consider global Q-balls in this scenario.  (As above, by global Q-balls we mean neglected gauged self-interactions.)  The VEV and energy per unit charge $\omega_0$ are shown in the left plot of Fig.~\ref{fig:BSM_global_omega_vev}, and the ratio of the energy per unit charge to the mass of a Higgs quanta is shown on the right.  These again are very similar to the Standard Model values, for the reasons outlined above.  Additional, we have studied the thin wall range of validity like we did for the Standard Model; the contribution of surface tension is negligible for global Higgs balls with charges $\gtrsim 10^5$.

\begin{figure}
    \centering
    \includegraphics[scale=0.61]{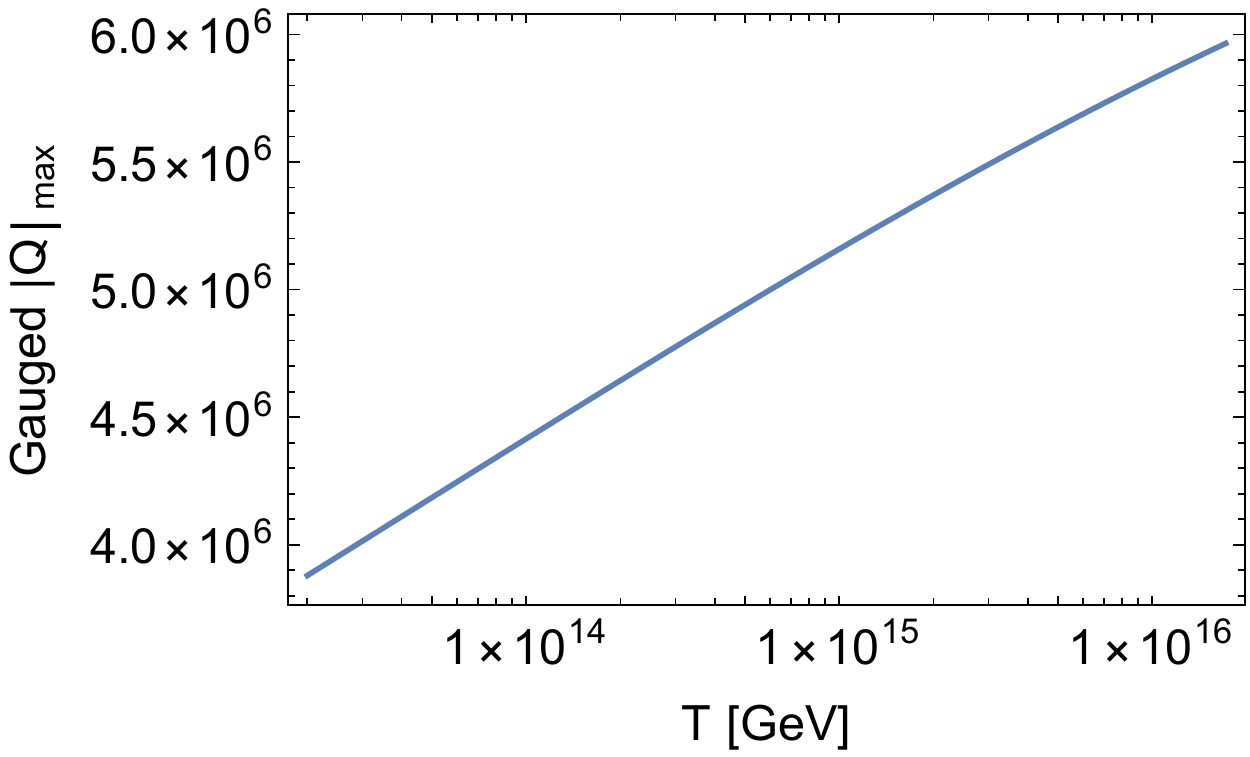}
    \includegraphics[scale=0.57]{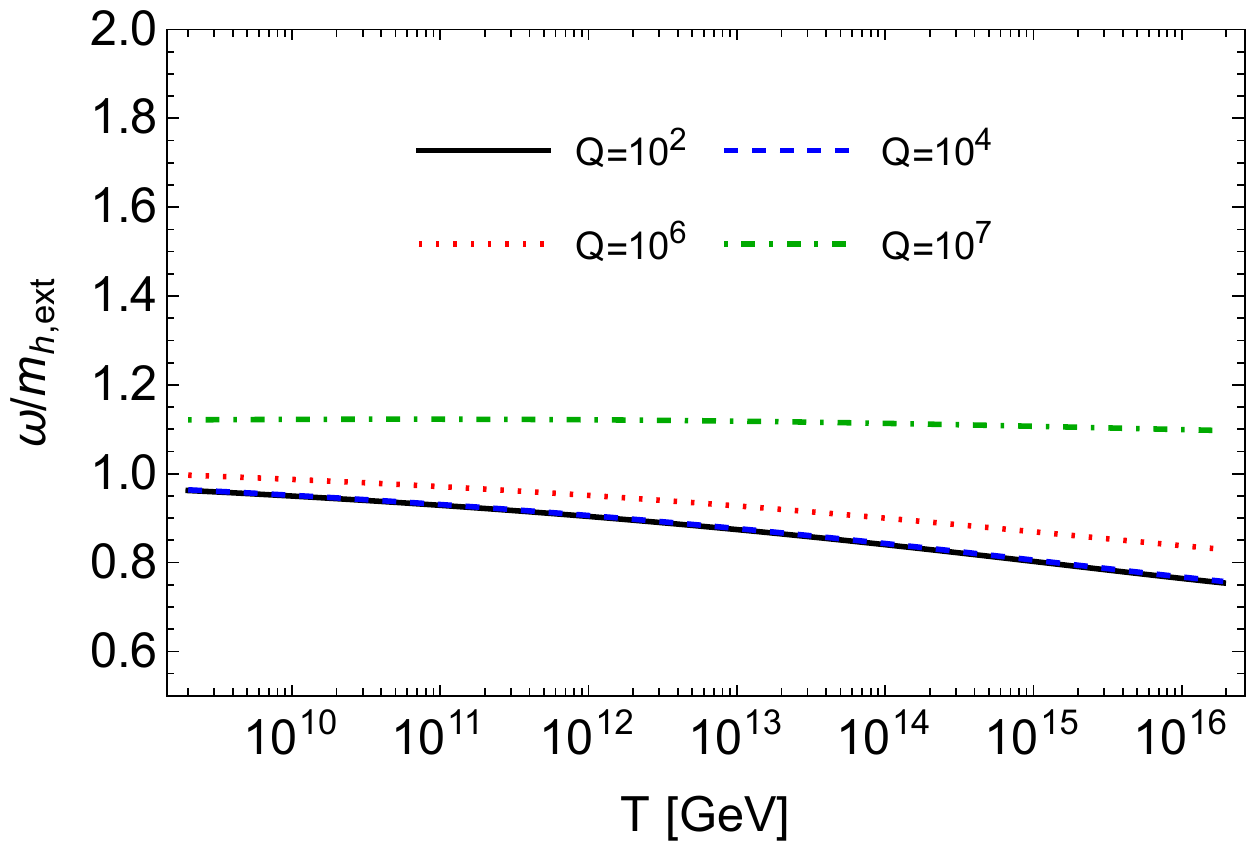}    
    \caption{Left: The maximum charge for which a gauged thin-wall Q-balls is stable, in our BSM scenario, as a function of temperature.  Right: Ratio of energy per unit charge to Higgs mass outside the Q-ball, as a function of temperature, for a range of charges.}
    \label{fig:BSM_gauged}
\end{figure}

\paragraph{}Thus, the energy of these Higgs balls (and their energy difference compared to an equivalent number of free Higgs quanta) is not significantly different from the pure Standard Model scenario.  However, because we have decreased the strength of the hypercharge and weak gauge couplings by a factor of 100, the destabilizing gauge self-interactions are significantly weaker.  We can use equations \eqref{eq:gauged_omega} and \eqref{eq:gauged_charge} to determine the energy per unit charge of a thin-wall gauged Higgs ball, and compare this to the mass of a Higgs quanta outside the Higgs ball.  This is shown in the plot of the left side of Fig.~\ref{fig:BSM_gauged}.  We see that Higgs balls with charges of order $\mathcal{O}(10^6)$ are stable across our temperature range.  On the right side, we present the analog of Fig.~\ref{fig:gauged_unstable} in our BSM scenario.  As expected from the left figure, we see that across the temperature range of interest, Higgs balls with charges up to $\mathcal{O}(6)$ are stable.  

\paragraph{}Furthermore, while surface effects have not been studied for gauged Q-balls, we expect the thin wall regime to be valid for these.  We note that global Q-balls with these charges are indeed in the thin wall regime, as mentioned above, and repulsive gauge self-interactions generally increase the size of the Q-balls, pushing them towards the thin wall regime.

\paragraph{}We note that we also studied a model in which the gauge couplings $g_Y$ and $g_W$ were modified to their running Standard Model values, divided by a factor of 10.  In this case, the maximum charge of stable Higgs balls was around $5000$.  As we were not confident these would be in the thin wall regime, we instead will focus on this benchmark scenario, in which the couplings are modified at high scales to be their running Standard Model values, divided by a factor of 100.

\section{Decay Rate of Higgs Balls}
\label{sec:decay}

\paragraph{}We have ensured that the Higgs balls discussed exist; that is, that they are stable against decay into individual Higgs quanta.  However, they are not absolutely stable as inside the Q-ball, where the Higgs VEV is nonzero, the Yukawa couplings allow Higgs quanta to decay into fermions (and  to photons via fermionic loops).  Additionally, the gauge couplings allow the Higgs quanta to decay into vector bosons.  These decays of individual quanta can lower the energy of the Higgs ball, and ultimately lead the overall soliton to decay.

\paragraph{}First, let us discuss new decays which may appear in our BSM mode.In the specific BSM scenario outlined above, the scalar singlet does not carry gauge charge and so the Higgs ball cannot decay into scalar quanta.  We also ensure the inside the Higgs ball, where $\mathrm{SU}(2) \times \mathrm{U}(1)$ is broken, decays to the scalar and its associated fermion are kinematically forbidden.  As noted above, we have assumed that the fermions which modify the running of the gauge couplings do not directly couple to the Higgs boson, and so any decays to these are loop suppressed (and furthermore, those loops involve the now-small gauge couplings).

\paragraph{}Therefore, in this section, we calculate the decay rate of the Q-balls using Standard Model decays.   In alternative BSM scenarios, new decay channels may be opened and in such cases, the rates considered here should be considered a lower bound on the decay rates.

\paragraph{}In Appendix~\ref{ap:decay_calcs}, we present the rates for a single Higgs quanta inside the Q-ball to decay into gauge bosons, photons, gluons, and fermions.  Bosonic decays can occur throughout the Q-ball, and therefore the Higgs ball's rate of decay into bosons is given by
\begin{align}
\Gamma_{\mathrm{Q-ball} \rightarrow \mathrm{bosons}} = Q \Gamma_{h \rightarrow \mathrm{bosons}}.
\end{align}

\paragraph{}It has been noted that frequently fermionic decays occur only at the surface of a Q-ball~\cite{Cohen:1986ct,Hong:2017uhi}, due to one of two factors~\cite{White:2021hwi}: First, decays can be kinematically forbidden due to the large VEV inside the Q-ball, which makes the fermions heavy.  However, our energy per unit charge is only about one order of magnitude smaller than the Higgs VEV, and therefore only the top quark decay is kinematically forbidden.

\begin{figure}
\centering
\includegraphics[scale=0.61]{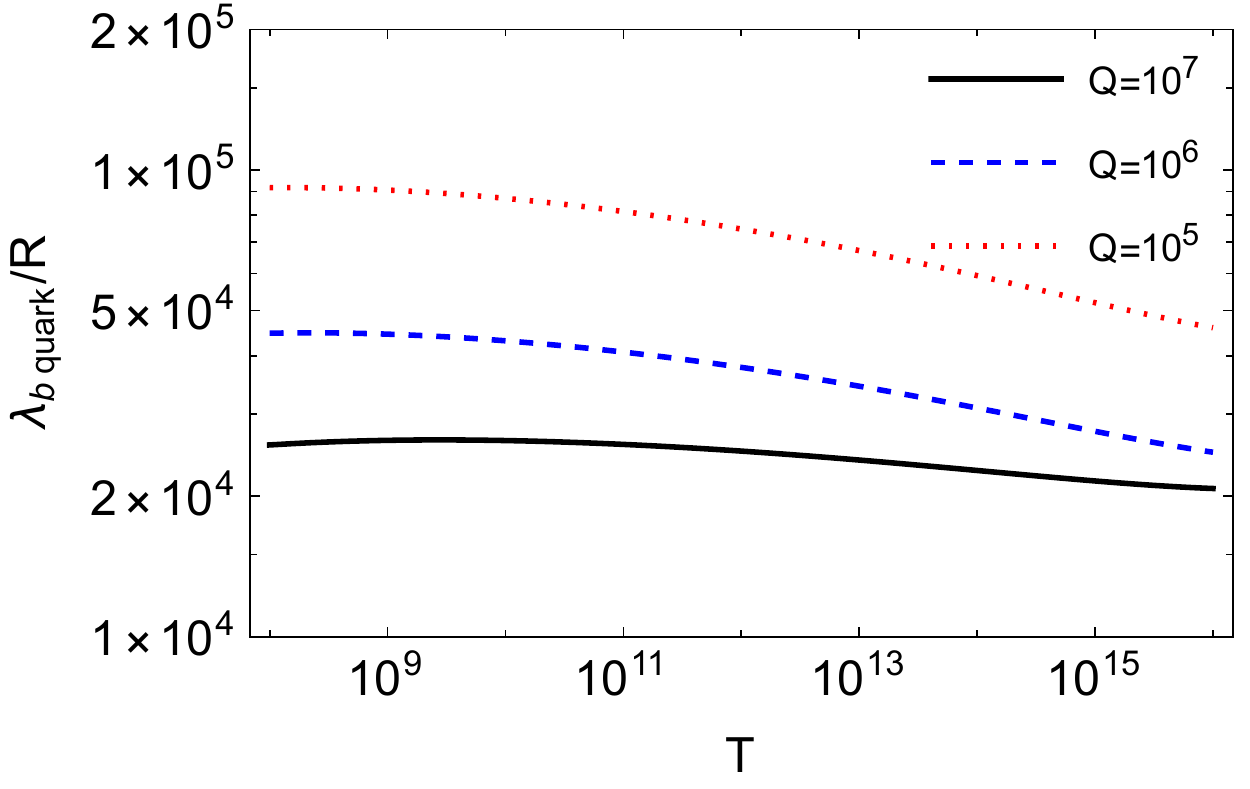}
\caption{The ratio of the mean free path of a b-quark produced via decay to the radius of the Higgs ball, for a variety of charges.  We observe that the quark can efficiently diffuse out of the Higgs ball.}
\label{fig:b_quark_mfp}
\end{figure}

\paragraph{}Secondly, inside the Q-ball the Fermi sea can fill up with the produced decay products, preventing further fermionic decays.  This occurs only if the decay products cannot diffuse out efficiently.  The mean free path is $\lambda \sim 1/{\sigma _{\psi \phi} n}$, where number density $n = 3 Q_0/ (4 \pi R^3)$ refers to the density of Higgs quanta inside the Q-ball.  To calculate the cross section, we consider scatterings via the gauge interactions and the Yukawa couplings, and approximate the momentum as $\omega \slash 2$.  Thus the cross section is taken to be $\sigma \sim \mathrm{Max} \left( g_Y^4 \slash (\omega \slash 2)^2, g_W^4 \slash (\omega \slash 2)^2, y_i^4 \slash (\omega \slash 2)^2 \right) $.  Comparing the mean free path to the radius of the gauged Higgs ball, we see that all the fermions are able to efficiently diffuse out; results for the b-quark are shown in Fig.~\ref{fig:b_quark_mfp}.  Therefore, fermionic decays also occur throughout the volume of the Higgs ball, giving a total decay rate for the overall Higgs ball of 
\begin{align}
\Gamma_{\mathrm{Q-ball} \rightarrow f \bar{f}} = Q \Gamma_{h \rightarrow f \bar{f}}.
\end{align}

\paragraph{}The production of Higgs balls is beyond this paper, as the production of gauged Q-balls is a complicated and not yet well-understood process.  However, as thermal Q-balls, Higgs balls require a large temperature, and a natural environment for them would be the early universe.  Therefore, we will consider their lifetime in this environment.  Cosmologically, Q-balls decay efficiently once the decay rate per unit charge is larger than the Hubble parameter, and therefore we consider 
\begin{align}
\dfrac{\Gamma_{\mathrm{Q-ball}}}{Q H } = \dfrac{\Gamma_h}{H},
\end{align}
since all of our decays occur throughout the volume of the Q-ball.  The only dependence on $Q$ is through the frequency of the gauged Q-balls, which varies weakly with charge as shown in the right plot of Fig.~\ref{fig:BSM_gauged}.  As shown on this plot, the variation between $Q = 10^6$ and $Q = 100$ (assuming the thin wall approximation is valid) is only a few percent, and therefore the results presented below are to high degree of precision independent of charge (although they are temperature dependent).  

This ratio is shown in Fig.~\ref{fig:higgs_ball_decay}, along with the net gauge boson and net fermionic contribution.  As explained in the Appendix, the gauge boson decay rate is dominated by longitudinal modes, while the fermion decay rate is dominated by b quarks.  The photon and gluon decay rates are highly suppressed.  We see that if they are produced in the early universe, Higgs balls are generally long lived at temperatures above about $2.6 \times 10^{13} \, \mathrm{GeV}$.

\begin{figure}
\centering
\includegraphics[scale=0.61]{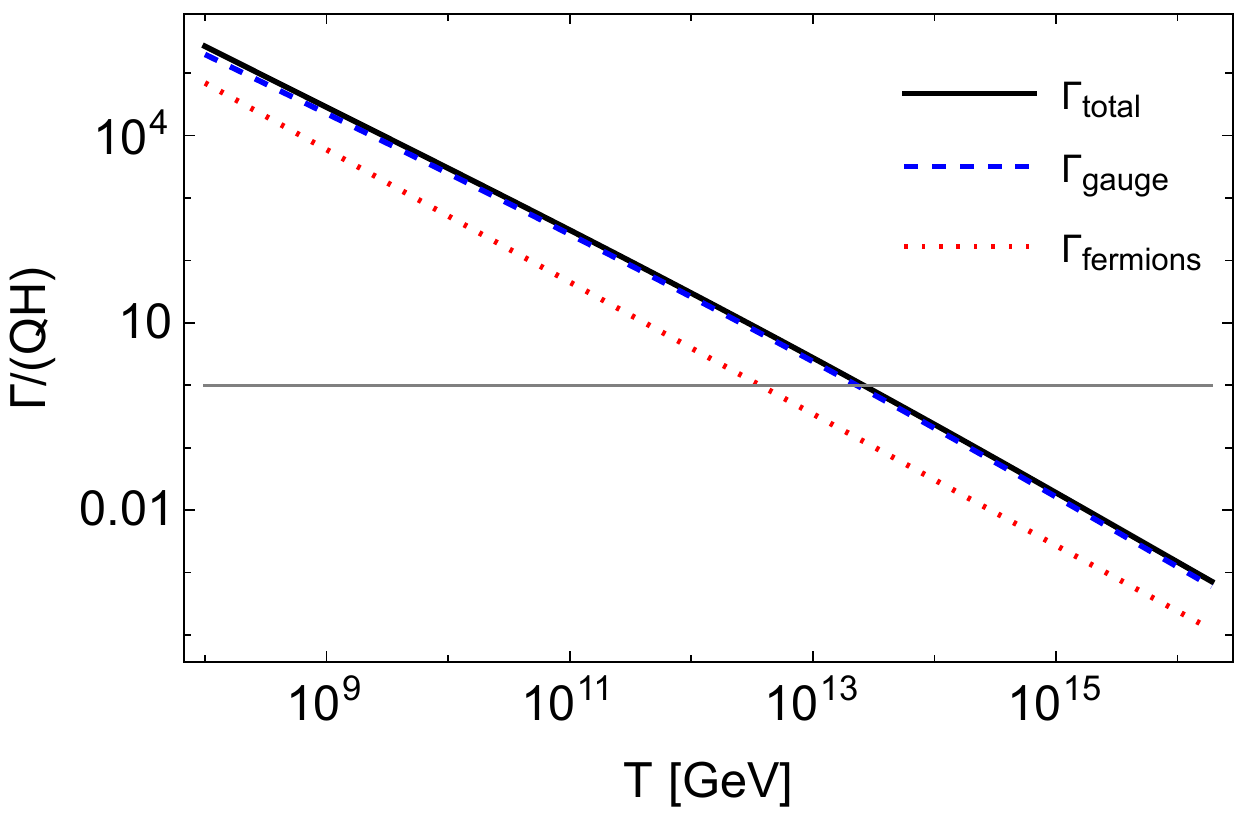}
\caption{The ratio of the decay rate of a Higgs ball to the Hubble parameter times charge, assuming radiation domination (for $Q = 10^6$).  Q-balls decay efficiently when this is larger than 1, which is indicated by the horizontal line.}
\label{fig:higgs_ball_decay}
\end{figure}

\paragraph{}As mentioned in the introduction, this can be usefully thought of in terms of an approximate conservation of Higgs number.  Because the Higgs field is the only scalar field that carries $\mathrm{SU}(2)$ and $\mathrm{U}(1)$ charge, to change the charge carried by the Q-ball the number of Higgs quanta must change.  As long as these processes are slow compared to the Hubble rate, there is approximate conservation of Higgs number charge, and the Higgs balls are long-lived.  Once these decays become efficient, this is no longer a conserved charge and the Higgs balls decay.  (We note that the decay of the Higgs quanta does not violate $\mathrm{SU}(2)$ and $\mathrm{U}(1)$ conservation, although the charge of the Q-ball changes; the gauge charge is carried out of the Q-ball by the decay products.)

\paragraph{}Since the gauge charge must be carried by scalar fields to produce a soliton, conserving gauge charge inside the Q-ball will generally be connected to keeping some sum of scalar field quanta constant.  Therefore, an approximate global symmetry can be usefully defined, which can be used to study stability, as we did here with Higgs number.

\section{Discussion and Conclusion}
\label{sec:conc}

\paragraph{}In this work, we have demonstrated the possibility of thermal balls, non-topological solitons that exist due to finite temperature corrections, even in models which have no attractive interactions at zero temperature and thus no Q-ball solutions at zero temperature.  The generic condition for the existence of such thermal balls is bosonic fields whose mass is determined by the scalar VEV.  

\paragraph{}If the bosonic field is a scalar, then the mere existence of a term proportional to $- A T |\phi|^3$ at high temperatures ensures the existence of these states, unless the scale at which these terms become relevant is unphysical (e.g., above the Planck scale).

\paragraph{}If the bosonic field is a gauge boson, then it's necessary to consider gauge effects, which can prevent such thermal balls from existing.  We have shown that this occurs in the Standard Model Higgs sector.  

\paragraph{}To show that thermal balls made of Higgs bosons (Higgs balls) can exist in extensions of the Standard Model, we have considered a model with two modifications: first, to weaken the gauge self-interactions, we have modified the running of the gauge couplings so that they are smaller at high scales.  Secondly, we have introduced an additional scalar field, which contributes to the thermal corrections as explained above.  Even with gauge effects included, Higgs balls exist in this model.

\paragraph{}We noted that Higgs balls won't be absolutely stable; even though they cannot decay into a set of Higgs quanta, they can decay into fermions and gauge bosons.  While the production of gauged Q-balls in the early universe has not been studied, we have shown that if they are produced, then for sufficiently high temperatures they will be long-lived. Of course, in other models, the thermal balls may be completely stable if no decays are permitted.

\paragraph{}We generally leave potential phenomenological implications for future work, as we have not discussed production of thermal balls.  However, we will note one intriguing connection.  Thermal balls require the existence of large cubic terms from thermal corrections.  If the model undergoes spontaneous symmetry breaking and these terms are relevant at those scales, then the phase transition will be first order~\cite{Buchmuller:1993bq}.  In the minimal modification of the Standard Model here, we have been agnostic about the scale of new physics; the existence of Higgs balls does not necessarily ensure a first order electroweak phase transition.  However, a first order electroweak phase transition, as envisioned in many models of baryogenesis \cite{Turok:1991uc,Funakubo:1993jg,McDonald:1993ey,Profumo:2006yu,Profumo:2014opa,Kozaczuk:2014kva,Vaskonen:2016yiu,Chiang:2017nmu,Baldes:2018emh,Kainulainen:2019kyp,Brdar:2019fur,Athron:2019teq,Huang:2020bbe,Niemi:2021qvp} and other well motivated scenarios involving phase transitions \cite{Croon:2018kqn,Croon:2018erz,Croon:2019rqu,Greljo:2019xan,Dev:2019njv,Fornal:2020ngq,Fornal:2021ovz}, does require significant cubic terms and hence we would expect thermal Higgs balls to exits in such models.



\paragraph*{}We also note that if produced in the early universe, Higgs balls could provide a novel means of dark matter production.  As Fig.~\ref{fig:BSM_global_omega_vev} demontrates, as the temperature changes the Higgs VEV also changes, and therefore the effective mass of the Higgs quanta inside the Higgs ball similar changes.  We consider models in which a new annihilation channel exists, which has a resonance at a specific Higgs mass (see for instance \cite{Davoudiasl:2004be,Patt:2006fw,GAMBIT:2018eea,Beniwal:2019xop,Arcadi:2019lka}).  As the Higgs mass evolves with temperature it will can pass through this resonance and thus produce dark matter \cite{Croon:2020ntf}.  However, since Higgs balls are only a fraction of the the volume, production will be suppressed.  This is desirable in e.g.\ Wimpzilla scenarios, in which one must avoid overproducing dark matter and overclosing the universe.  However, specific abundances cannot be calculated without knowing the Higgs ball production rate which we leave to future work. 

\paragraph{Acknowledgements:} The authors thank Daniel Vagie and Volodymyr Takhistov for helpful conversations.  The works of AK and GW were supported by World Premier International Research Center Initiative (WPI), MEXT, Japan. A.K. was supported the U.S. Department of Energy (DOE) grant No. DE-SC0009937 and by Japan Society for the Promotion of Science (JSPS) KAKENHI grant No.
JP20H05853. 

\bibliographystyle{JHEP}
\bibliography{references}

\appendix

\section{Standard Model Potential}
\label{ap:SM_potential}

\paragraph{}In this appendix, we discuss how we calculate the Standard Model potential at zero and finite temperature.  Our conventions and resummations follow~\cite{Arnold:1992rz}.   We use the Landau gauge, following~\cite{Quiros:1994dr,DelleRose:2015bpo}.  The renormalization group improved tree potential is
\begin{equation}
V_{\rm tree} (h) = - \dfrac{\mu(h)^2}{2} h^2 + \dfrac{\lambda(h)}{4} h^4,
\end{equation}
where, inside the H-ball, we set our renormalization scale to be the Higgs VEV.  We use SARAH~\cite{Staub:2013tta} to evaluate our running coupling constants, which evaluates the renormalization group equations to two-loop order.

\paragraph{}We initialize the Standard Model parameters in SARAH at the Z pole, following the procedure in~\cite{Croon:2020cgk}, although with all parameters related to the singlet set to zero.  In particular, in matching the $\overline{\mathrm{MS}}$ parameters to observed pole masses, we include one loop corrections for the Standard Model Higgs parameters $\mu$ and $\lambda$, the top quark Yukawa $y_t$, and the electroweak gauge coupling $g$.  All other parameters are matched at tree-level.  This is explained further in Appendix \ref{ap:matching}.

\paragraph{}In the potential, we also include the one-loop corrections~\cite{Quiros:1994dr,DelleRose:2015bpo}
\begin{equation}
V_{1-{\rm loop}}(h) = \sum_{W,Z,t,\chi,h} \dfrac{n_i}{64 \pi^2} m_i(h)^4 \left[ \ln \left(\dfrac{m_i(h)^2}{h^2} \right)- C_i \right]
\end{equation}
(where the contributions of all other fields are subdominant).  In this expression, we again take the renormalization scale to be the Higgs VEV; the coefficients are:
\begin{align}
n_W=6, n_Z=3, n_t = -12, n_\chi = 3, n_h = 1, \nonumber \\
C_W=C_Z = 5 \slash 6, C_t = C_\chi = C_h = 3 \slash 2.
\end{align}
The masses are
\begin{align}
m_W(h)^2 &= \dfrac{1}{4} g_2(h)^2 h^2, \nonumber \\
m_Z(h)^2 &= \dfrac{1}{4} (g_1(h)^2 + g_2(h)^2) h^2, \nonumber \\
m_t(h)^2 &= \dfrac{1}{2} y_t(h)^2 h^2, \nonumber \\
m_\chi(h)^2 &= - \dfrac{\mu(h)^2}{2} + \lambda(h) h^2, \nonumber \\
m_h(h)^2 &= - \dfrac{\mu(h)^2}{2} + 3 \lambda(h) h^2 ,
\end{align}
with all running parameters evaluated at the Higgs VEV inside the H-ball.  (We note that at the scales of interest, $- \mu(h)^2 \ll \lambda(h) h^2$ and this contribution to the Higgs and $\chi$ masses can be neglected.)

\paragraph{}As noted in Section~\ref{sec:SM_Hballs}, finite temperature corrections are important to H-ball existence.  We include the one-loop finite temperature correction, 
\begin{align}
V_{1-\mathrm{loop}}(h,T) 
&= 
	\sum_{\mathrm{bosons}} \dfrac{n_iT^4}{2 \pi^2} J_B \left( \dfrac{m_i^2}{T^2} \right) 
	+ \sum_{\mathrm{fermions}} \dfrac{n_i T^4}{2 \pi^2} J_F \left( \dfrac{ m_i^2}{T^2} \right),
\end{align}
where 
\begin{align}
	J_B(y) &= \int_0^\infty dx \, x^2 \ln \left( 1 - e^{- \sqrt{x^2+y} }\right) , \nonumber \\
	J_F(y) &= \int_0^\infty dx \, x^2 \ln \left( 1 + e^{- \sqrt{x^2+y} }\right) ,
\end{align}
Again we neglect contributions from fermions other than the top, as they are highly suppressed.  (We note that the constant term proportional to $T^4$, which does not depend on the Higgs VEV, has no impact on our solitons and therefore can be neglected.)

\paragraph{}We also include the so-called daisy diagrams via
\begin{align}
&V_{\rm ring}(h,T) = \sum_{W_L,Z_L,\gamma_L,\chi,h} \dfrac{n_i T^4}{12 \pi} \left( \left( \dfrac{m_i^2}{T^2} \right)^{3 \slash 2} - \left( \dfrac{\mathcal{M}_i^2}{T^2} \right)^{3 \slash 2} \right) ,
\end{align}
where the Debye masses for $h$, $\chi$, and the longitudinal mode of the $W$ boson are given by $\mathcal{M}_i^2(h) = m_i^2(h) + \Pi_i(h,T)$ with
\begin{align}
\Pi_h(h,T) &= \Pi_\chi(h,T) = \left( \dfrac{3 g_2^2 + g_1^2}{16} + \dfrac{\lambda}{2} + \dfrac{y_t^2}{4} \right) T^2 \nonumber \\
\Pi_{W_L}(h,T) &= \dfrac{11}{6} g_2^2 T^2.
\end{align}
Due to the mixing between the $Z$ boson and the photon, their Debye masses are
\begin{align}
\mathcal{M}^2_{Z_L}(h) &= \dfrac{1}{2} \left( m_Z^2 + \dfrac{11}{6} \dfrac{g_2^2 T^2 }{\cos^2(\theta_W)} + \Delta(h,T) \right) \nonumber \\
\mathcal{M}^2_{Z_L}(h) &= \dfrac{1}{2} \left( m_Z^2 + \dfrac{11}{6} \dfrac{g_2^2 T^2 }{\cos^2(\theta_W)} - \Delta(h,T) \right) ,
\end{align}
where
\begin{align}
\Delta^2(h,T) &= m_Z^4 + \dfrac{11}{3} \dfrac{g^2 \cos^2(\theta_W)}{\cos^2(\theta_W)}  \cdot \left( 
m_Z^2 + \dfrac{11}{12} \dfrac{g_2^2 T^2}{\cos^2(\theta_W)}
\right) T^2.
\end{align}
In this expression, all couplings (including the Weinberg angle) are running couplings, evaluated at a renormalization scale set by the Higgs vacuum expectation value.  As above, we have also subtracted off the $h$-independent terms proportional to $T^4$, which do not impact our solitons.

\paragraph{}In our analysis we have integrated $J_B$ and $J_F$ numerically; however, we note that in the high temperature limit,
\begin{align}
V(h,T) &= V_{1-\mathrm{loop}}(h,T) + V_{\rm ring}(h,T)  \nonumber \\
&\approx \kappa T^2 h^2  - AT h^3,
\end{align}
where 
\begin{align}
\kappa &= \frac{1}{32} \left(g_1^2+3 g_2^2+8 \lambda +4 y_t^2\right) -\frac{ \lambda}{6 \pi}  \sqrt{g_1^2+3 g_2^2+8 \lambda +4 y_t^4}+\frac{1}{32 \pi } \sqrt{\dfrac{11}{6}} (g_1^3 + 3 g_2^3) , \nonumber \\
A &= \frac{8 \left(\sqrt{3}+1\right) \lambda^{3/2} + \left(g_1 ^2+ g_2^2\right)^{3/2}+2 g_2^3}{32 \pi },
\end{align}
which illustrates the $-A T h^3$ term that allows for Higgs balls to exist.

\paragraph{}In discussions of the stability of the H-balls, we have compared the energy of the H-ball to the energy of $Q$ free Higgs quanta.  For this, we used the thermal mass of the Higgs boson in the true vacuum, given by $m_h(T) = \sqrt{ 2 \kappa} $.  In the true vacuum (where the Higgs VEV is zero) we have evaluated the running couplings at a renormalization scale set by the temperature.

\section{Matching Standard Model Parameters to Observables}
\label{ap:matching}

\paragraph{}As mentioned in section \ref{sec:SM_Hballs}, we want to avoid Q-balls that induce solitosynthesis, which means that we want to ensure the Standard Model vacuum remains stable.  The stability of the Standard Model vacuum is sensitive to the precise values of measured parameters such as the top mass, and therefore one must take care when matching observables to the running Standard Model parameters.  We describe our procedure here.

\paragraph{}Following Ref.~\cite{Croon:2020cgk}, the one loop self energy for the top quark can be written as
\begin{equation}
    \frac{1}{2}y_t^2 v^2 = M_t^2\left( 1+ 2 {\rm Re} \left(  \Sigma _v (M_t^2) + \Sigma _s (M_t^2) \right) \right) ,
\end{equation}
where
\begin{align}
\Sigma_v(M^2_t) + \Sigma_s(M^2_t) &=
    \frac{3}{16} \frac{g^2_2}{(4\pi)^2} \bigg( -2 - 4 \frac{1}{h^2} - 2 h^2 - \frac{256}{9}s^2 + 2 t^2 + 16 \frac{t^4}{h^2}
    \nonumber \\ &
    - \frac{2}{27} \Big( 39 - \frac{64}{z^2} + 25 z^2 + 18\frac{z^4-1}{h^2} \Big)
    \nonumber \\ &
    + \Big(4 h^2 - \frac{8}{3}t^2 + \frac{4}{3} t^2 \frac{2t^2 + h^2}{t^2-h^2} \Big) \ln(h) - \frac{8}{9} \Big( -9 \frac{z^4}{h^2} + 4 \frac{(4-5z^2+z^4)}{t^2-z^2} \Big) \ln(z)
    \nonumber \\ &
    + \Big( \frac{128}{3}s^2 -32 \frac{t^4}{h^2} - \frac{4}{3} \frac{t^2(2t^2 + h^2)}{t^2-h^2} - \frac{32}{9} \frac{(z^2-1)(t^2-4) }{z^2-t^2} \Big) \ln(t)
    \nonumber \\ &
    + \frac{2}{3}\Big(4t^2 -h^2  \Big) F(M_t,M_t,M_h)
    + \frac{2}{3}\frac{(t^2+2)(t^2-1)}{t^2} F(M_t,M_W,0)
    \nonumber \\ &
    - \frac{2}{27}\Big( \frac{64 - 80z^2 + 7z^4}{z^2}
     + \frac{32-40z^2 + 17z^4}{t^2} \Big) F(M_t,M_t,M_Z)
    \nonumber \\ &
    + \bigg[ 2\Big(-6 \frac{1}{h^2} - h^2 -\frac{32}{3}s^2 + t^2 + 8 \frac{t^4}{h^2} \Big)
    \nonumber \\ &
    - \frac{4}{9}\frac{(z^2-1)(9+4h^2+9z^2)}{h^2} \bigg] \ln \Big( \frac{Q^2}{M^2_W} \Big) \bigg)
    \nonumber \\ &
    - 64 c_6 \frac{M^2_W}{g^4_0} \bigg[1 - \ln \Big( \frac{Q^2}{M^2_h} \Big) \bigg] \bigg)
    \ ,\\[3mm]
\frac{\delta g^{2}}{g^2_0} &=
    \frac{1}{(4\pi)^2} g^2_0  \bigg(
    - \frac{257}{72}
    - \frac{1}{24}h^2
    + \frac{20}{9} N_F
    + \frac{1}{4}t^2 
    - 2\ln(t)
    \nonumber \\ &
    + \frac{1}{12}(12 - 4 h^2 + h^4) F(M_W,M_h,M_W)
    - \frac{(t^2+2)(t^2-1)}{2} F(M_W,M_t,0)
    \nonumber \\ &
    - \frac{33}{4}F(M_W,M_W,M_W)
    + \Big(\frac{4}{3}N_F -\frac{43}{6} \Big) \ln \Big( \frac{Q^2}{M^2_W} \Big)
    \bigg),
\end{align}
and where the mass ratios are $\{ h,t,z\} =\{ M_h/M_W, M_t/M_W, M_Z/M_W \}$. (In this section, we use $v$ for the Higgs VEV.)  The $\overline{\mathrm{MS}}$ RG scale is denoted by $Q$, and $F(k,m_1,m_2)$ is a loop function given in Ref.~\cite{Kajantie:1995dw}.  We adjusted the top mass pole mass away from its central value of $172.4 \pm 0.7$ GeV~\cite{ParticleDataGroup:2020ssz} as explained in section \ref{sec:SM_Hballs} to stabilize the vacuum up to the desired scale.

For the Higgs sector parameters we again use the self energies of the W and Higgs bosons,
\begin{eqnarray}
\mu _h ^2 &=& - \frac{1}{2} M_h ^2 \left( 1+ \frac{{\rm Re}\Pi _h (M_h ^2)}{M_h ^2}\right) \\
\lambda &=& \frac{1}{8} g_2^2 \frac{M_h^2}{M_W^2} \left(1 + \frac{{\rm Re} \Pi _h (M_h^2)}{M_h^2} - \frac{{\rm Re} \Pi _W (M_W^2)}{M_W^2} \right)
\end{eqnarray}
where to one loop, the self energies have the form
\begin{align}
\Pi_h(M^2_h) &=
    \frac{3}{8} \frac{g^2_0 M^2_h}{(4\pi)^2} \bigg(
    - \frac{4}{3}
    - 8 \frac{1}{h^2}
    - 2h^2
    + 16 \frac{t^4}{h^2}
    - \frac{2}{3} z^2
    - 4\frac{z^4}{h^2}
    \nonumber \\ &
    + 3h^2 F(M_h,M_h,M_h)
    + 4t^2\Big( 1 - 4 \frac{t^2}{h^2} \Big) F(M_h,M_t,M_t)
    \nonumber \\ &
    + \frac{2}{3}\frac{h^4 - 4h^2 + 12}{h^2} F(M_h,M_W,M_W)
    \nonumber \\ &
    + \Big( \frac{1}{3} \frac{1}{h^2} - \frac{4}{3}z^2 + 4\frac{z^4}{h^2} \Big) F(M_h,M_Z,M_Z)
    \nonumber \\ &
    - 2h^2 \ln(h)
    - 8t^2 \ln(t)
    + \Big( -\frac{2}{3}h^2 + 4 z^2 \Big) \ln(z)
    \nonumber \\ &
    + \Big( -4 +2h^2 + 4 t^2 - 2 z^2 \Big) \ln\Big( \frac{Q^2}{M^2_W} \Big)
    \nonumber \\ &
    + 64 c_6 \frac{M^2_W}{g^4_0 h^4} \bigg[ -2 + 12 t^4 - z^4 +  3 h^4 F(M_h,M_h,M_h)
    - 6 h^4 \ln(h)
    \nonumber \\ &
    - 24 t^4 \ln(t)
    + 6z^4 \ln(z)
    + \Big( -2 + h^4 + 4 t^4 - z^4 \Big) \ln\Big( \frac{Q^2}{M^2_W} \Big) \bigg]
    \nonumber \\ &
    + 3072 c_6^{2} \frac{M^4_W}{g^4_0 h^2} \bigg[ -1 + F(M_h,M_h,M_h) \bigg]\bigg)
    \;,
\end{align}
and
\begin{align}
\Pi_W(M^2_W) &=
    \frac{3}{8} \frac{g^2_0 M^2_W}{(4\pi)^2} \bigg(
    - \frac{212}{9}
    - \frac{8}{3}\frac{1}{h^2}
    - \frac{22}{9} h^2
    + \frac{4}{27} (40N_F - 17)
    - \frac{4}{3}t^2
    + 16 \frac{t^4}{h^2}
    + \frac{14}{9}z^2
    - \frac{4}{3} \frac{z^4}{h^2}
    \nonumber \\ &
    + \frac{4h^2 (h^2-2)}{h^2-1} \ln(h)
    - 8\Big( \frac{2}{3} - t^2 + 4\frac{t^4}{h^2} \Big) \ln(t)
    + 4\Big(2 \frac{z^4}{h^2} - \frac{z^4 - 4 z^2 - 8}{z^2-1} \Big) \ln(z)
    \nonumber \\ &
    + \frac{2}{9}\Big( 12 - 4 h^2 + h^4 \Big) F(M_W,M_h,M_W)
    - \frac{4}{3}(t^2+2)(t^2-1) F(M_W,M_t,0)
    \nonumber \\ &
    - \frac{32}{3}\frac{z^2-1}{z^2} F(M_W,M_W,0)
    + \frac{2}{9}\frac{(z^4 + 20z^2 + 12)(z^2-4)}{z^2} F(M_W,M_W,M_Z)
    \nonumber \\ &
    + 2\bigg[-1 + \frac{2}{h^2} + \Big( -\frac{59}{9} - 6 \frac{1}{h^2} - h^2 + \frac{16}{9}N_F - 2 t^2 + 8 \frac{t^4}{h^2} \Big) + z^2 - 2\frac{z^4}{h^2} \bigg] \ln\Big( \frac{Q^2}{M^2_W} \Big)
    \nonumber \\ &
    - \frac{8}{9}\pi i (4N_F - 3)
    - 64 c_6 \frac{M^2_W}{g^4_0} \bigg[1 - \ln\Big( \frac{Q^2}{M^2_h} \Big) \bigg]  \bigg)
    \;. 
\end{align}

\section{Gauged Q-balls in the Static Charge Approximation}
\label{ap:gauging_qballs}

\paragraph{}In this appendix, we consider gauged Q-balls in the static charge approximation following Ref.~\cite{Heeck:2021zvk}, which studies gauged $\mathrm{U}(1)$ Q-balls.  We first consider a gauged $\mathrm{SU}(N)$ scenario before discussing the Standard Model Higgs, which carries $\mathrm{SU}(2) \times \mathrm{U}(1)$ charge.

\paragraph{}As we show, the static charge approximation does not include the effects of gauge boson self-interactions, which generally occur in non-Abelian gauge theories.  Therefore it is valid when the size of the Q-ball is less than the confinement scale.  All unbroken $\mathrm{SU}(N)$ interactions other than $\mathrm{U}(1)$ confine on some scale, and therefore it is expected that the static charge approximation breaks down.  We note that a Q-ball which carries $\mathrm{SU}(N)$ charge will necessarily be screened by charges exterior to the Q-ball at the confinement scale. To estimate the $\mathrm{SU}(2)$ confinement scale we run $g_2$ using the one loop RGE
\begin{eqnarray}
\mu \frac{d \alpha _W}{d\mu} = - \frac{b_W}{2 \pi }\alpha _W
\end{eqnarray}
with 
\begin{equation}
    b_W = \frac{22}{3}-\frac{n_F}{3} - \frac{n_s}{6} \ .
\end{equation}
As a result we find that the $\mathrm{SU}(2)$ confinement scale is at $\mu = O(10^{-10}) ({\rm GeV})$.  This gives a confinement scale $R_{\rm conf} \sim 1 \slash \mu$ which is significantly larger than the radii of our Q-balls throughout the temperature ranges considered in this work.  We can therefore ignore macroscopic confinement effects. 
\paragraph{}We start with a complex scalar field $\phi$ which transforms under a gauged $\mathrm{SU}(N)$, with generators $t^a$ and structure constant $f^{abc}$.  The relevant Lagrangian density is 
\begin{align}
\mathcal{L} &= |D_\mu \phi|^2 - U(|\phi|) - \dfrac{1}{4} F^{a\dagger}_{\mu \nu} F^{a \mu \nu},
\end{align}
where the field strength tensor is
\begin{align}
F^a_{\mu \nu} &= \partial_\mu A_\nu^a - \partial_\nu A_\mu^a + g f^{abc} A_\mu^b A_\nu^c
\end{align}
and the gauged covariant derivative is
\begin{align}
D_\mu &= \partial_\mu - i g A_\mu^a t^a.
\end{align}

\paragraph{}We introduce a generalized static charge ansatz 
\begin{align}
\phi(t,\vec{x}) &= \dfrac{1}{\sqrt{2}} \phi_0 F(r) e^{i \omega t},
\end{align}
where $F(r)$ is real, although $\phi_0$ can be complex.  In this ansatz, the gauge fields are
\begin{align}
A_0^a(t,\vec{x}) &= A_0^a(r), \qquad A_a^i(t,\vec{x}) = 0,
\end{align}
where $A_0^a(r)$ will be taken to be real.

\paragraph{}After simplifying, we have the Lagrangian
\begin{align}
L &= (4\pi) \int dr \, r^2 \left[ 
\dfrac{1}{2}  \phi_0^\dagger \phi_0 \left(  \omega^2 F(r)^2 - (F^\prime(r))^2 \right)   
-  \omega g  A_0^a(r)  \phi_0^\dagger t^a \phi_0 F(r)^2 
\right. \nonumber \\
& \qquad  \left. 
+ \dfrac{g^2}{4} A^{a}_0 A_0^b F(r)^2 \phi_0^\dagger  \left\lbrace t_a,t_b \right\rbrace \phi_0 
- U\left( \dfrac{1}{2} \phi_0^\dagger \phi_0 F(r)^2 \right) 
+\dfrac{1}{2} (A_0^{a\prime}(r))^2  
\right] .
\end{align}

\paragraph{}We now specialize to $\mathrm{SU}(2)$ and assume that $\phi_0$ transforms in the fundamental representation.  Without a loss of generality we take the VEV to be 
\begin{align}
\phi_0 &= \begin{pmatrix}
	 0 \\ v
\end{pmatrix},
\label{eq:phi_0}
\end{align}
giving 
\begin{align}
L &= (4\pi) \int dr \, r^2 \left[ 
\dfrac{v^2}{2}  \left(  \omega^2 F(r)^2 - (F^\prime(r))^2 \right)   
+  \dfrac{\omega g}{2}  A_0^3(r)  v^2 F(r)^2 
+ \dfrac{g^2 v^2}{8} A^{a}_0 A_0^a F(r)^2 
\right. \nonumber \\
& \qquad  \left. 
- U\left( \dfrac{v^2}{2}  F(r)^2 \right) 
+\dfrac{1}{2} (A_0^{a\prime}(r))^2  
\right] ,
\end{align}
where our choice of the VEV direction selects the $A^3$ gauge field.

\paragraph{}Next, we introduce rescalings like those in Ref.~\cite{Heeck:2021zvk}.  In particular, we define:
\begin{alignat}{4}
\Phi_0 &= \dfrac{v_0}{\sqrt{m_\phi^2 - \omega_0^2}}, \qquad   & 
\rho &= r \sqrt{ m_\phi^2 - \omega_0^2}, \qquad   &
\Omega &= \dfrac{\omega}{\sqrt{m_\phi^2 - \omega_0^2}},  \nonumber \\
A^a(\rho) &= \dfrac{A_0^a(\rho)}{v},\qquad   &
\alpha &= g \Phi_0, \qquad &  \kappa^2 &= \Omega^2 - \Omega_0^2 ,
\qquad   & 
\Omega_0 &= \dfrac{\omega_0}{\sqrt{m_\phi^2 - \omega_0^2}},
\end{alignat}
and now we will use a prime to indicate a derivative with respect to $\rho$.  The Lagrangian is then 
\begin{align}
L 
&= 4\pi \Phi_0^2\sqrt{m_\phi^2 - \omega_0^2}  \int d\rho \, \rho^2  \left[ 
- \dfrac{1}{2}   (F^\prime(\rho))^2 
+ \dfrac{1}{2}  (A^{a\prime}(\rho ))^2   
+ \dfrac{1}{2} F^2 \left(  \Omega^2
+  \Omega \alpha   A^3(\rho) 
+  \dfrac{ \alpha^2 }{4} (A^{a})^2  \right) 
\right. \nonumber \\
& \qquad  \left. 
- \dfrac{U\left( F(\rho)^2 \right) }{\Phi_0^2 (m_\phi^2 - \omega_0^2)^2 }
\right] 
\end{align}
from which we find the equations of motion
\begin{align}
0 &=  F^{\prime \prime} + \dfrac{ 2}{ \rho} F^\prime  + F \left(  \Omega^2
+  \Omega \alpha   A^3(\rho) 
+  \dfrac{ \alpha^2 }{4} (A^{a})^2  \right)
-  \dfrac{1}{\Phi_0^2 (m_\phi^2 - \omega_0^2)^2 } \dfrac{dU}{df} \nonumber \\
0 &= (A^{3})^{\prime \prime} + \dfrac{2}{\rho} (A^3)^\prime  
- \dfrac{\alpha}{2} F^2 \left( \Omega +  \dfrac{\alpha}{2} A^3 \right) \nonumber \\ 
0 &= (A^{1})^{\prime \prime} + \dfrac{2}{\rho} (A^1)^\prime  
	- \dfrac{\alpha}{2} F^2 \left(  \dfrac{\alpha}{2} A^1 \right) \nonumber \\
0 &= (A^{2})^{\prime \prime} + \dfrac{2}{\rho} (A^2)^\prime  
- \dfrac{\alpha}{2} F^2 \left(  \dfrac{\alpha}{2} A^2 \right) .
\end{align}

\paragraph{}Assuming the scalar field carries unit charge, the charge of the Q-ball, $Q = -i\int d^3x \, (\Phi^\dagger D_0 \Phi - h.c.)$, is given by 
\begin{align}
Q &= 4 \pi
\int d\rho \, \rho^2 F(\rho)^2 \Phi_0^2
\left( \Omega 
+ \dfrac{\alpha}{2} A^3(\rho) \right) ,
\end{align}
which after using the equations of motion and an integration by parts, is 
\begin{align}
Q &= \dfrac{4\pi \Phi_0^3}{\alpha \slash 2} \lim_{\rho \rightarrow \infty} \rho^2 A_3^\prime,
\end{align}
similar to $\mathrm{U}(1)$ gauged Q-balls.  In particular, we see that \begin{align}
\lim_{\rho \rightarrow \infty} A_3 = -\dfrac{\alpha Q\slash 2}{4 \pi \Phi_0^2 \rho }.
\end{align}

\paragraph{}We observe that regardless of the function $F(\rho)$, $A^1 = A^2 = 0$ is a solution to the equations of motion, and so therefore we set these fields to zero inside the Q-ball.  The resulting equations for $F$ and $A_0^3$ are identical to those in Ref.~\cite{Heeck:2021zvk} with $\alpha \rightarrow - \alpha \slash 2$.  Following this reference, we similarly impose the boundary conditions
\begin{align}
\lim_{\rho \rightarrow 0} F^\prime 
= \lim_{\rho \rightarrow \infty} F 
= \lim_{\rho \rightarrow 0} A_3^\prime 
= \lim_{\rho \rightarrow \infty} A_3 = 0
\end{align}

Therefore, if we approximate the scalar field profile with a step function $F(\rho) = 1 - \Theta(\rho - R^*)$, and impose $A^3$ and $(A^3)^\prime$ are continuous at the boundary, we find 
\begin{align}
A^3(\rho) &= -\dfrac{2 \Omega}{\alpha } \begin{cases}
1 - \dfrac{\sinh(\alpha \rho \slash 2)}{\cosh(\alpha R^* \slash 2)(\alpha \rho \slash 2)}, \qquad&
 \rho < R^* \\
\dfrac{  \alpha R^* \slash 2 - \tanh(\alpha R^* \slash 2)}{\alpha \rho \slash 2}, \qquad &
\rho \geq R^*.
\end{cases}
\end{align}

\paragraph{}The derivative of $\alpha A^3$ is small if the radius is large, and therefore in the equation of motion for $F$ we can approximate $A^3$ with its value at the radius.  Then we recognize that the equation for $F$ is that of a global Q-ball, but with the frequency given by
\begin{align}
\Omega_G&= \Omega + \dfrac{\alpha}{2} A^3(R^*),
\end{align}
giving a relation between the energy per unit charge of the gauged Q-ball and the energy per unit charge of the ungauged Q-ball,
\begin{align}
\Omega &= \Omega_G \cdot \dfrac{\alpha R^*}{2} \coth \left( \dfrac{\alpha R^*}{2} \right) .
\end{align}
In the thin wall limit, this becomes
\begin{align}
\omega &= \omega_0 \cdot \dfrac{\alpha R^*}{2} \coth \left( \dfrac{\alpha R^*}{2} \right) .
\end{align}

\paragraph{}Now we turn our attention to the Standard Model Higgs, which carries both $\mathrm{SU}(2)$ and $\mathrm{U}(1)$ charge.  We start with the Lagrangian density
\begin{align}
\mathcal{L} &= |D_\mu \phi|^2 - U(|\phi|) - \dfrac{1}{4} G^{a\dagger}_{\mu \nu} G^{a \mu \nu}
- \dfrac{1}{4} F^{\dagger}_{\mu \nu} F^{ \mu \nu},
\end{align}
where the field strength tensors are
\begin{align}
G^a_{\mu \nu} &= \partial_\mu A_\nu^a - \partial_\nu A_\mu^a + g_W f^{abc} A_\mu^b A_\nu^c \nonumber \\
F_{\mu \nu} &= \partial_\mu B_\nu - \partial_\nu B_\mu
\end{align}
and the gauged covariant derivative is
\begin{align}
D_\mu &= \partial_\mu - i g_W A_\mu^a t^a - \dfrac{i}{2} g_Y B_\mu.
\end{align}
We use the convention that the Higgs has hypercharge $1 \slash 2$.  Proceeding as above, and making the static charge ansatz for both the $\mathrm{SU}(2)$ gauge field $A\mu^a$ and the $\mathrm{U}(1)$ gauge field $B_\mu$, we find the Lagrangian
\begin{align}
L 
&= 4\pi \Phi_0^2\sqrt{m_\phi^2 - \omega_0^2}  \int d\rho \, \rho^2  \left[ 
- \dfrac{1}{2}   (F^\prime(\rho))^2 
+ \dfrac{1}{2}  (A^{a\prime}(\rho ))^2 - \dfrac{U\left( F(\rho)^2 \right) }{\Phi_0^2 (m_\phi^2 - \omega_0^2)^2 }  \right. \nonumber \\
& \left. 
+ \dfrac{1}{2} F^2 \left(  \Omega^2
+  \Omega \alpha_W   A^3(\rho) 
+  \dfrac{ \alpha_W^2 }{4} (A^{a}(\rho))^2  
- \Omega \alpha_YB(\rho) + \dfrac{\alpha_Y^2}{4} (B(\rho))^2
- \dfrac{\alpha_W \alpha_Y}{2} B(\rho) A^3(\rho) 
\right) 
\right] .
\label{eq:lagrangian}
\end{align}
In this expression, $\alpha_W = g_W \Phi_0$, $\alpha_Y = g_Y \Phi_0$, $A^a(\rho) = A_0^a(\rho) \slash v$, and $B(\rho) = B_0(\rho) v$.

\paragraph{}We note that again the equation of motion for the radial $F(\rho)$ matches that of a  global Q-ball, with a shifted frequency,
\begin{align}
0 &=  F^{\prime \prime} + \dfrac{ 2}{ \rho} F^\prime  + F \left(  
\Omega + \dfrac{\alpha_W}{2} A^3 -  \dfrac{\alpha_Y}{2} B \right)^2
-  \dfrac{1}{\Phi_0^2 (m_\phi^2 - \omega_0^2)^2 } \dfrac{dU}{df} \nonumber \\
0 &= (A^{3})^{\prime \prime} + \dfrac{2}{\rho} (A^3)^\prime  
- \dfrac{\alpha_W}{2} F^2 \left( \Omega +  \dfrac{\alpha_W}{2} A^3 - \dfrac{ \alpha_Y}{2} B \right) , \nonumber \\
0 &= B^{\prime \prime} + \dfrac{2}{\rho} B^\prime  
+ \dfrac{\alpha_Y}{2} F^2 \left( \Omega -  \dfrac{\alpha_Y}{2} B^3 + \dfrac{\alpha_W}{2} A_3 \right) ,
\end{align}
and the other fields $A_1$ and $A_2$ may be set to zero.  We see that the equations of motion mix $A^3$ and $B$, as is expected for the Higgs sector.

A Q-ball made of the Higgs field will carry both $\mathrm{U}(1)$ and $\mathrm{SU}(2)$ charge.  We note that since we have chosen the VEV in the form above, the amount of $\mathrm{SU}(2)$ charge is described by the weak isospin.  A quanta of the Higgs field has hyperchange $1 \slash 2$ and weak isospin $-1 \slash 2$, so the charges of the Higgs ball satisfy $Q_Y = - Q_W = - \dfrac{i}{2} \int d^3x \, \left(  \Phi^\dagger D_0 \Phi - h.c. \right)$, as it must for it to be electrically neutral.  Using our equation of motion above, we see that this can be expressed using either gauge field
\begin{align}
	Q_Y = - Q_W  &= \dfrac{4\pi \Phi_0^2}{\alpha_W \slash 2} \lim_{\rho \rightarrow \infty} \rho^2 A_3^\prime
= - \dfrac{4\pi \Phi_0^2}{\alpha_B \slash 2} \lim_{\rho \rightarrow \infty} \rho^2 B^\prime,
\label{eq:charges}
\end{align}
which again determines their behavior at large radii.  

\paragraph{}Although the equations are coupled, they can be solved in the thin wall approximation.  Taking $F(\rho) = 1 - \Theta(\rho - R^*)$ and imposing the boundary conditions 
\begin{align}
\lim_{\rho \rightarrow 0} F^\prime 
= \lim_{\rho \rightarrow \infty} F 
= \lim_{\rho \rightarrow 0} A_3^\prime 
= \lim_{\rho \rightarrow \infty} A_3 = 0
= \lim_{\rho \rightarrow 0} B\prime 
= \lim_{\rho \rightarrow \infty} B = 0,
\end{align}
as well as imposing the continuity of $A^3$ and $B$ at $R^*$ (as well as their derivatives), gives
\begin{align}
A^3(\rho) &= \begin{cases} 2 \alpha_W \Omega \left( 
\frac{2 \sinh \left(\frac{1}{2} \rho  \sqrt{\alpha_W^2+\alpha_Y^2}\right) \text{sech}\left(\frac{1}{2} R^* \sqrt{\alpha_W^2+\alpha_Y^2}\right)}{\rho  \left(\alpha_W^2+\alpha_Y^2\right)^{3/2}}-\frac{1 }{\alpha_W^2+\alpha_Y^2} \right) &   \rho < R^* \nonumber \\
-\frac{2 \alpha_W \Omega  \left(R^* \sqrt{\alpha_W^2+\alpha_Y^2}-2 \tanh \left(\frac{1}{2} R^* \sqrt{\alpha_W^2+\alpha_Y^2}\right)\right)}{\rho  \left(\alpha_W^2+\alpha_Y^2\right)^{3/2}}&   \rho \geq R^*
\end{cases} 
\end{align}
and:
\begin{align}
B(\rho) = \begin{cases}
2 \alpha_Y \Omega  \left(\frac{1}{\alpha_W^2+\alpha_Y^2}-\frac{2 \sinh \left(\frac{1}{2} \rho  \sqrt{\alpha_W^2+\alpha_Y^2}\right) \text{sech}\left(\frac{1}{2} R^* \sqrt{\alpha_W^2+\alpha_Y^2}\right)}{\rho  \left(\alpha_W^2+\alpha_Y^2\right)^{3/2}}\right)  &   \rho < R^* \nonumber \\
 \frac{2 \alpha_Y \Omega  \left(R^* \sqrt{\alpha_W^2+\alpha_Y^2}-2 \tanh \left(\frac{1}{2} R^* \sqrt{\alpha_W^2+\alpha_Y^2}\right)\right)}{\rho  \left(\alpha_W^2+\alpha_Y^2\right)^{3/2}}
 & \rho \geq R^* 
\end{cases}
\end{align}

We note that in the large $\rho$ regime, $B(\rho) = - \dfrac{\alpha_Y}{\alpha_W} A^3(\rho)$, as required by $Q_Y = - Q_W$.  We remind the reader that in this expression, $\alpha_W$ and $\alpha_Y$ are the quantities defined under equation \eqref{eq:lagrangian}, and not the couplings squared over $4 \pi$.

From the equation for $F$ above, we observe that the energy per unit charge is related to that of a global Q-ball by
\begin{align}
	\Omega_G&= \Omega + \dfrac{\alpha_W}{2} A^3(R^*) -  \dfrac{\alpha_Y}{2} B(R^*),
\end{align}
which gives us 
\begin{align}
\omega =\frac{1}{2} R v_0 \omega_0 \sqrt{g_W^2+g_Y^2} \coth \left(\frac{1}{2} R v_0 \sqrt{g_W^2+g_Y^2}\right)
.
\end{align}

\paragraph{}Substituting the solutions into \eqref{eq:charges}, we find
\begin{align}
Q_Y = - Q_W &= \frac{8 \pi  R \omega_0 \left(R v_0 \sqrt{g_W^2+g_Y^2} \coth \left(\frac{1}{2} R v_0 \sqrt{g_W^2+g_Y^2}\right)-2\right)}{g_W^2+g_Y^2},
\end{align}
which scales as $R^3 \omega_0 v_0^2$ in the appropriate limit.  Replacing $v_0$ with $h_0$ for the Higgs VEV gives the expression in section \ref{sec:gauging}.  We also note that since we have used the step function ansatz, $F = 1$ inside the Higgs ball, and therefore the VEV is $h_0$, the same value as for the global Higgs ball.

\section{Modifications for our BSM Scenario}
\label{ap:BSM_Changes}

\paragraph{}In this appendix, we will summarize the changes we made in our BSM (Beyond-the-Standard-Model) scenario.  We assume that somewhere between the scales probed by current experiments and the lowest scales relevant (temperature and/or Higgs VEV of order $\mathcal{O}(10^8)$) new fermions affect the running of the Standard Model gauge couplings, suppressing them.  Since we are interested in the general phenomenon of Higgs balls, we we implemented this simply by dividing the running Standard Model couplings by a factor of 100.  

\paragraph{}The primary modification to the Higgs potential comes from the additional scalar $S$.  This scalar field generally helps to stabilize the Higgs potential through a positive contribution to the running quartic coupling.  The quartic term in the potential will typically dominate at scales slightly above the mass of the singlet.  Consequently, Q-balls would exist only in a narrow temperature regime around the mass of the singlet, and a broad study of thermal balls is difficult.  (This would also likely decrease any phenomenological impact they may have.)

\paragraph{}Therefore, we also introduce a fermion $\psi$ at the same scale as the scalar, with a coupling constant $y_f = \sqrt{2 \lambda_{HS}}$.  With this choice, the one-loop corrections to the Standard Model Higgs sector quartic coupling $\lambda$ from the scalar and fermion cancel with each other.  This could perhaps be justified via an embedding in a supersymmetric model, although we leave the for future study.  Therefore, we use the Standard Model running couplings (with $g_Y$ and $g_W$ modified as explained above).

\paragraph{}As mentioned above, we assume the fermions either do not couple to the Standard Model Higgs directly, or such couplings are suppressed so that their contribution to the Higgs potential is negligible.  

\paragraph{}As in Appendix \ref{ap:SM_potential}, we follow the approach in Ref.~\cite{Arnold:1992rz}.  As the scalar does not acquire a vacuum expectation value, $V_{\rm tree}(h)$ is unchanged.  The scalar and fermion contributions to $\Delta V_{1-\mathrm{loop}}$ exactly cancel, due to the relation between $\lambda_{HS}$ and the Yukawa coupling discussed above.

 \paragraph{}The contribution to the finite temperature  potential is 
\begin{align}
\Delta V_{1-\rm{loop}}(h,T) &= \dfrac{T^4}{2\pi^2} J_B \left( \dfrac{m_{S,\rm{eff}}^2}{T^2} \right)
- \dfrac{T^4}{2 \pi^2} J_F \left(\dfrac{m_f^2}{T^2} \right)
,
\end{align} 
where $m_{S,\rm{eff}} = \sqrt{ \lambda_{HS} h} $ and the mass of the fermion is $\sqrt{ \lambda_{HS}} h$.  We note that this is independent of the gauge couplings, as desired.

\paragraph{}Finally, the contribution to the daisy diagrams is
\begin{align}
\Delta V_{\rm{ring}}(h,T) &= \dfrac{T^4}{12 \pi} \left( \left( \dfrac{m_{S,\rm{eff}}^2}{T^2} \right)^{3 \slash 2} - 
\left( \dfrac{\mathcal{M}_S^2}{T^2} \right)^{3 \slash 2} \right),
\end{align}
where the Debye mass is $\mathcal{M}_S^2 = m_{S,\rm{eff}}^2 + \Pi_S(h,T)$ and
\begin{align}
\Pi_S(h,T) = \dfrac{\lambda_{HS}}{12} T^2.
\end{align}
As before, we have taken care to subtract off terms proportional $T^4$ which are independent of the VEV $h$ are therefore do not affect our solitons.  Finally, we note that the Debye masses of the Higgs and Goldstone bosons $\chi$ are also modified by the scalar and fermion.  We add to these
\begin{align}
\Delta \Pi^2_{h,\chi} = \left( \dfrac{\lambda_{HS}}{12}+ \dfrac{y_f^2}{4} \right) T^2.
\end{align}

\section{Decay Rates of Higgs Quanta Inside the Higgs Ball}
\label{ap:decay_calcs}

\paragraph{}In this appendix, we describe our calculations for the decay rate of Higgs quanta inside the Higgs ball into Standard Model particles, which is used in section \ref{sec:decay} to calculate the decay rate of the Higgs ball.  At tree-level, the Higgs quanta inside the Higgs ball can decay directly to gauge bosons and fermions.  In addition to considering these, we note that at loop level it can decay to photons and gluons.  We also consider these decays, and show that they are suppressed.

\paragraph{}Before proceeding, we note the following: In appendix \ref{ap:gauging_qballs} we showed that the Higgs VEV inside the gauged Q-balls is the same as in the global Q-ball, and so we evaluate the decay rates at $h = h_0$.  We also made sure to evaluate these at the energy per unit of the gauged, not global, Higgs balls.

\paragraph{Gauge Boson Decay:} As noted in Appendix \ref{ap:SM_potential}, the longitudinal and transverse modes of the gauge bosons receive different corrections to their masses at finite temperature.  Consequently, we consider their decay rates separately.  The decay rates for a single Higgs quanta inside the Higgs ball to decay to gauge bosons are given by
\begin{align}
\Gamma_{h \rightarrow WW,\mathrm{trans}} 
&= \dfrac{g_W^4 h^2}{32 \pi}  \cdot  \dfrac{ \sqrt{ 1 - 4 x_{W,\mathrm{trans}} }}{\omega},	\nonumber \\
\Gamma_{h \rightarrow ZZ,\mathrm{trans}} 
&= \dfrac{1}{2} \dfrac{(g_W^2 + g_Y^2)^2 h^2}{32 \pi}  \cdot  \dfrac{ \sqrt{ 1 - 4 x_{Z,\mathrm{trans}} }}{\omega},	\nonumber \\
\Gamma_{h \rightarrow WW,\mathrm{long}} &= \dfrac{1}{256 \pi}  \cdot  \dfrac{g_W^4 h^2 \omega^3}{m_{W,\mathrm{long}}^4}  \cdot \sqrt{1 - 4 x_{W,\mathrm{long}}} \left( 1 - 2 x_{W,\mathrm{long}} \right)^2,  \nonumber \\
\Gamma_{h \rightarrow ZZ,\mathrm{long}} &= \dfrac{1}{2} \cdot \dfrac{1}{256 \pi}  \cdot  \dfrac{(g_W^2 + g_Y^2)^2 h^2 \omega^3}{m_{Z,\mathrm{long}}^4}  \cdot \sqrt{1 - 4 x_{Z,\mathrm{long}}} \left( 1 - 2 x_{Z,\mathrm{long}} \right)^2  ,
\end{align}
where $x_i = M_i^2 \slash \omega^2$, where inside the Higgs ball the effective mass of a single Higgs quanta is $\omega$, the energy per unit charge.  We note that at zero temperature, in which $m_{W,\mathrm{long}} = g_W h \slash 2$ and $m_{Z,\mathrm{long}} = \sqrt{ g_W^2 + g_Y^2} h \slash 2$, the longitudinal decay rates have a piece independent of the gauge coupling, as required by the Golstone Boson Equivalence Theorem.  Using these relations and summing over the modes reproduces the standard result (e.g.,~\cite{Spira:2016ztx}), provided that one also replaces $\omega$ with the Higgs mass.

\paragraph{}Because $g_W$ and $g_Y$ are suppressed in our BSM model, the decays to the transverse modes are also highly suppressed.  However, the decays to the longitudinal modes are not, and it turns our these will be the dominant decay modes of the quanta inside the Higgs ball.  

\paragraph{Photon Decay:}  The decay of the Higgs quanta to photons (and gluons) is suppressed by loop effects.  At zero temperature, the photon decay rate is given by~\cite{Spira:2016ztx}
\begin{equation}
\Gamma(h \rightarrow \gamma \gamma )
= \dfrac{G_F \alpha^2 m_h^3}{128 \sqrt{2} \pi^3 }
\bigg| \sum_f N_{cf} e_f^2 A_f^H(\tau_f)
+A_W^H(\tau_W) \bigg|^2,
\end{equation}
where
\begin{align}
	A_F^H(\tau) &= 2 \tau \left( 1 + (1-\tau) f(\tau) \right), \nonumber \\
	A_W^H(\tau) &= - \left( 2 + 3 \tau + 3 \tau (2-\tau) f(\tau) \right),
\end{align}
and $\tau_i = 4 M_i^2 \slash m_h^2$.  $N_{cf}$ is one for leptons and three for quarks, and $e_f$ is the charge of each species in units of the proton charge.  In these relations, the function $f$ is given by
\begin{align}
	f(\tau) = \begin{cases}
		\arcsin(1 \slash \sqrt{\tau})^2 & \tau \geq 1, \nonumber \\
		- \dfrac{1}{4} \left[ \log \left( \dfrac{1 + \sqrt{ 1 - \tau}}{1 - \sqrt{ 1 - \tau}} \right) - i \pi \right]^2 & \tau < 1.	
	\end{cases}
\end{align}

\paragraph{}The first term, from fermion loops, can be easily generalized to decays of Higgs quanta inside Q-balls at finite temperature, by replacing $m_H$ with $\omega$ and using $G_F = 1 \slash (\sqrt{2} h^2)$.  However, the second term, from $W$ boson loops, is more complicated.  At finite temperature, it should be evaluated in the Landau gauge, and since the transverse and longitudinal parts of the gauge boson receive different Debye masses, these should be separately identified; for example, by using a gauge boson propagator which projects out the longitudinal and transverse components~\cite{Buchmuller:1993bq}.

\paragraph{}As we expect the photon contribution to be negligible due to loop suppression, and the smallness of $\alpha$, we will not carefully calculate it but estimate it as follows.  First, we write $G_F = g_W^4 h^2 \slash (\sqrt{2} m_W^4)$, motivated by the fact that the $hWW$ vertex introduces a factor of $g_W^2$ into the matrix element.  At zero temperature, when $m_W = g h \slash 2$, this reduces to the familiar $1 \slash (\sqrt{2} h^2)$.  Secondly, we note that the first term in $A_W^H$, which is independent of the $W$ boson mass, must come from the longitudinal mode by the Goldstone Boson Equivalence Theorem, as this gives the contribution which is independent of the gauge coupling.  The remaining pieces of $A_W^H$ are a sum of the longitudinal and transverse contributions.

\paragraph{}We will thus overestimate the decay rate if we calculate:
\begin{equation}
\Gamma(h \rightarrow \gamma \gamma )
\leq \dfrac{\alpha^2 \omega^3}{256 \pi^3 }
\bigg| \dfrac{1}{h} \sum_f N_{cf} e_f^2 A_f^H(\tau_f)
+\dfrac{g_W^2 h}{m_{W,\mathrm{long}}^2} 
A_W^H(\tau_{W,\mathrm{long}}) 
+\dfrac{g_W^2 h}{m_{W,\mathrm{trans}}^2} 
\tilde A_W^H(\tau_{W,\mathrm{trans}}) 
\bigg|^2,
\end{equation}
where $\tilde{A}_W^H$ simply leaves off the first term of $A_W^H$, and we also replace $m_H$ with $\omega$ when calculating $\tau$ (in addition to including thermal masses).  Using this, we have verified that $\Gamma_{h \rightarrow \gamma \gamma}$ is many orders of magnitude smaller than the gauge boson decay rate throughout the temperature range of interest. 

\paragraph{Gluon Decay:} In general, the Higgs quanta may also decay to gluons via fermion loop effects.  The leading order decay rate is~\cite{Spira:2016ztx} 
\begin{align}
	\Gamma(h \rightarrow gg) &=
	\dfrac{ \alpha_s^2 \omega^3}{72 v^2  \pi^3} 
	\bigg| \sum_Q A_Q^H(\tau_Q) \bigg|^2,
\end{align}
where
\begin{align}
	A_Q^H(\tau) &= \dfrac{3}{2} \tau \left( 1 + (1-\tau) f(\tau) \right)
\end{align}
and the function $f(\tau)$ was given above.  The sum is over all the quarks, using thermally corrected mass.  However, the gluon also has an induced thermal mass~\cite{Comelli:1997ru}, which are
\begin{align}
m_{G,\mathrm{long}} &= g_S^2 T \nonumber \\
m_{G,\mathrm{trans}} &= \sqrt{ 2 g_S^2 T^2 } + \dfrac{3}{4\pi} g_S^2 T \ln \left( \dfrac{\sqrt{ 3 + n_Q \slash 2}}{g_S} \right),
\end{align}
where $n_Q$ is the number of quarks in the thermal plasma.  These thermal masses are sufficient to make the decay kinematically forbidden throughout the region of interest, due to the smallness of the energy per unit charge $\omega$.

\paragraph{Fermionic Decays:} Next we turn our attention to fermionic decays.  As noted in Section \ref{sec:decay}, fermionic decays occur throughout the Higgs ball as the decay products are able to efficiently diffuse out.  The rate for a single Higgs quanta to decay into fermions is
\begin{align}
\Gamma_{h \rightarrow f \bar{f}}
&= \dfrac{3 \omega y_f^2}{8 \pi}, 
\end{align}
where we use the running Yukawa couplings.  The Higgs can decay to all fermions except the top quark, although we neglect neutrino decays as they are highly suppressed.  As expected, decays to bottom quarks dominate due to the larger Yukawa coupling.

\begin{figure}
    \centering
    \includegraphics[scale=0.6]{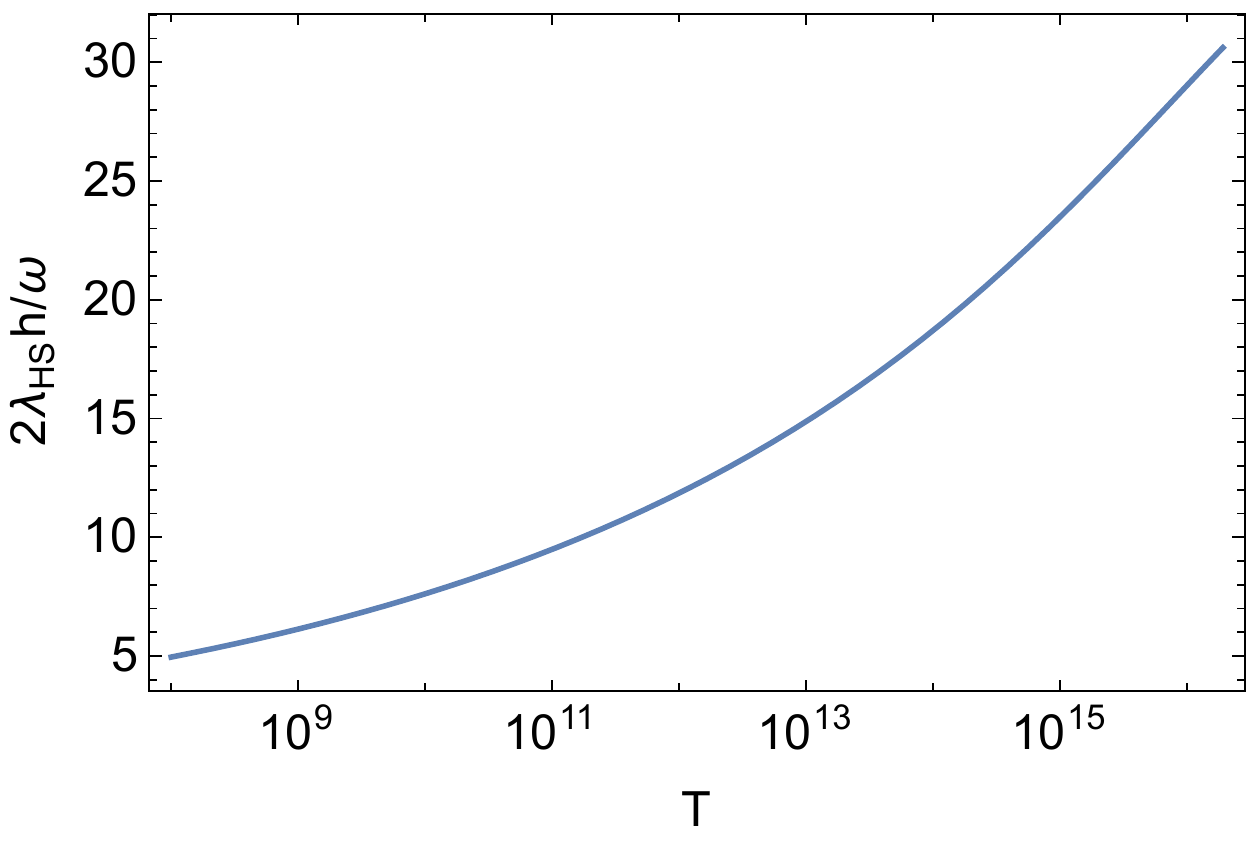}
    \caption{The ratio of $2 m_{S,\rm{eff}} = 2 m_\psi = 2 \sqrt{ \lambda_{HS}} h$ to the energy per unit charge inside the Higgs ball.  Since this is greater than one, decays into the scalar $S$ and fermion $\psi$ are kinematically forbidden.  This is evaluated at $Q = 10^6$, but it depends on the charge very weakly for reasons explained in Section \ref{sec:decay}.}
    \label{fig:No_BSM_Decays}
\end{figure}

\paragraph{BSM Decays:} Our BSM model introduced several new states: a set of fermions which alter the running of the $g_Y$ and $g_W$ couplings, a scalar singlet $S$, and a fermion $\Psi$ whose role is to cancel the singlet's contribution to the running Higgs quartic coupling.  All of these could introduce new decay channels for the Higgs.

\paragraph{}Since we are primarily interested in long-lived Higgs balls, we want to suppress these decays.  Therefore, we do not couple the fermions which alter the gauge running to the Higgs boson, or restrict their Yukawa couplings to be highly suppressed so that these decays are negligible.

\paragraph{}The large value $\lambda_{HS} = 0.9$ induces a large mass for the singlet, and since $y_f = \sqrt{2 \lambda_{HS}}$, for the fermion also.  In particular, $m_{S,\rm{eff}} = m_\psi = \sqrt{\lambda_{HS}} h$.  In Fig.~\ref{fig:No_BSM_Decays}, we show the ratio $2 \sqrt{\lambda_{HS}} h \slash \omega$, which shows that it is kinematically forbidden in the range of interest.

\end{document}